\documentclass[preprint,5p,times]{elsarticle}

\usepackage{graphicx}
\usepackage{bm}
\usepackage{caption}
\usepackage{subcaption}
\usepackage{tabularx}
\usepackage{algorithm}
\usepackage{algpseudocode}
\usepackage{listings}
\usepackage{hyperref}
\usepackage[bb=boondox]{mathalfa}
\usepackage{amsmath}   
\usepackage{amssymb}
\usepackage{upgreek}   
\usepackage{xcolor}
\usepackage{booktabs}
\usepackage{multirow}
\usepackage{makecell}  
\usepackage[normalem]{ulem}  
 
\newcommand\myPi{\mathrel{\stackrel{\makebox[0pt]{\mbox{\normalfont\tiny B}}}{\pi}}}

\newcommand{\myBigPi}{\makebox{\LARGE\ensuremath{\pi}}}
\newcommand{\myBigPiB}{\makebox{\LARGE\ensuremath{\myPi}}}

\newcommand{\myBigSigma}{\makebox{\Large\ensuremath{\sigma}}}

\newcommand{\myBigGamma}{\makebox{\Large\ensuremath{\upgamma}}}
\newcommand{\myGamma}{\makebox{\large\ensuremath{\upgamma}}}
\newcommand\myCup{\mathrel{\stackrel{\makebox[0pt]{\mbox{\normalfont\tiny B}}}{\cup}}}
\newcommand{\myBigUnionB}{\makebox{\Large\ensuremath{\myCup}}}
\newcommand{\myBigCup}{\makebox{\large\ensuremath{\cup}}}
\newcommand{\myBigJoin}{\makebox{\large\ensuremath{\bowtie}}}

\newcommand{\myPiB}{\makebox{\large\ensuremath{\myPi}}}
\newcommand{\myUnionB}{\makebox{\large\ensuremath{\myCup}}}

\newcommand{\myBigRho}{\makebox{\large\ensuremath{\rho}}}

\newcommand{\myBigAlpha}{\makebox{\large\ensuremath{\alpha}}}
\newcommand{\myMediumPi}{\makebox{\Large\ensuremath{\pi}}}

\newcommand{\mySmallSigma}{\makebox{\Large\ensuremath{\sigma}}}

\newcommand{\myBigLeftSemijoin}{\makebox{\Large\ensuremath{\ltimes}}}

\newcommand{\myBigRightSemijoin}{\makebox{\Large\ensuremath{\rtimes}}}

\newcolumntype{Z}{>{\raggedright\arraybackslash}X} 
\newcolumntype{Y}{>{\centering\arraybackslash}X}
\newcolumntype{R}{>{\raggedleft\arraybackslash}X}

\begin{document}

\begin{frontmatter}



\title{A DBMS-independent approach for capturing provenance polynomials through query rewriting}


\author[first]{Paulo Pintor}
\author[second]{Rog\'erio Lu\'is de Carvalho Costa}
\author[first]{Jos\'e Moreira}

\affiliation[first]{
  organization={University of Aveiro, Department of Electronics, Telecommunications and Informatics},
  addressline={University Campus Santiago}, 
  city={Aveiro},
  postcode={3810-193}, 
  state={},
  country={Portugal}
}

\affiliation[second]{
  organization={CIIC, Polytechnic of Leiria},
  addressline={School of Technology and Management}, 
  city={Leiria},
  postcode={2411-901}, 
  state={},
  country={Portugal}
}

\begin{abstract}

In today's data-driven ecosystems, ensuring data integrity, traceability and accountability is important. Provenance polynomials constitute a powerful formalism for tracing the origin and the derivations made to produce database query results. Despite their theoretical expressiveness, current implementations have limitations in handling aggregations and nested queries, and some of them and tightly coupled to a single Database Management System (DBMS), hindering interoperability and broader applicability.

This paper presents a query rewriting-based approach  for annotating Structured Query Language (SQL) queries with provenance polynomials. The proposed methods are DBMS-independent and support Select-Projection-Join-Union-Aggregation (SPJUA) operations and nested queries, through recursive propagation of provenance annotations.
This constitutes the first full implementation of semiring-based theory for provenance polynomials extended with semimodule structures. It also presents an experimental evaluation to assess the validity of the proposed methods and compare the performance against state-of-the-art systems using benchmark data and queries. The results indicate that our solution delivers a comprehensive implementation of the theoretical formalisms proposed in the literature, and demonstrates improved performance and scalability, outperforming existing methods.
\end{abstract}



\end{frontmatter}



\section{Introduction}
\label{sec:introduction}

Data provenance is an important field in information systems as it captures detailed metadata about the origin and derivation processes of database query results~\cite{Seattle2022,Singh2019}. 
This enables improved 1) data lineage and traceability, by documenting the data lifecycle, from the origin to the final output, 2) reproducibility, by capturing how data was processed and which operations were applied, 3)  debugging, by helping to trace back through intermediate steps to locate sources of error, whether in the input data or the derivations, among other applications~\cite{Herschel2017,Fernandez2020}. 
This is increasingly  important in the context of modern data ecosystems, where data comes frequently from multiple sources and is accessed through distributed query engines that unify access across relational, NoSQL and other data sources, using a single query interface~\cite{Seattle2022,Shen2019,Barclay2019}.

Different types of data provenance~\cite{Buneman2001,where_provenance,Herschel2017} and formalisms are covered in the literature. A notable case is the provenance semiring~\cite{Green2007}, an algebraic structure used to model the derivation of tuples in the result of database queries. In~\cite{Green2017}, provenance semirings are organised in a hierarchy with varying levels of expressiveness to capture data provenance information. It ranges from simple structures like lineage semirings to complex structures named provenance polynomials. The latter is at the top of the hierarchy and provides a comprehensive symbolic representation of how query results are derived from input data. They are particularly relevant because the other types of provenance semirings in the hierarchy can be derived from provenance polynomials.

The structure and operators of provenance semirings are insufficient for capturing the information about aggregation functions such as \texttt{Sum}, \texttt{Count}, or \texttt{Max}. To address this limitation, it was proposed an extension of the semirings approach that incorporates monoids~\cite{Yael2011}. This extension introduces new operators and expressions that enable the representation of aggregation function results within the semirings framework.

Despite the well-established theoretical foundations in provenance semirings and provenance polynomials~\cite{Green2007,Yael2011,Green2017}, existing implementations only partially address existing challenges. Namely, the extension proposed in~\cite{Yael2011} has been implemented as a tightly coupled single DBMS solution (ProvSQL~\cite{Senellart2018}) that has limited functionality. It is unable to process nested queries and exhibits limitations when applying filters, such as \texttt{Having} clauses, in conjunction with aggregations. This limitation hinders the full potential of provenance polynomials in data analytics and decision-making environments. Perm~\cite{Glavic2009} is designed exclusively for PostgreSQL, like ProvSQL. 

GProM~\cite{Arab2018}, is a middleware built on Perm that enables access to various databases (PostgreSQL, Oracle, and SQLite) but lacks supporting  
distributed query engines. 
Another limitation in existing proposals involves obtaining provenance results for complex queries, such as nested queries requiring provenance propagation. For instance, when nested queries appear in the \texttt{Where} clause, existing solutions may demand significant query rewriting to perform unnesting and enable provenance propagation. In this regard, Perm is one of the few systems capable of addressing this challenge.

This paper introduces DBMS-independent methods that use provenance annotations and query rewriting techniques to capture provenance polynomials and facilitate annotation propagation in the context of nested queries. The annotations effectively represent provenance polynomials based on semiring theory~\cite{Green2007,Green2017}, incorporating the enhancement proposed in~\cite{Yael2011} to account for aggregation information. 

The proposed solution aligns with the SQL standard, eliminating the need for users to learn new operators, modify the original query, or establish additional support structures to obtain provenance results. It also supports distributed query engines (e.g. Trino~\cite{trino}) to execute queries in distributed environments and introduces a nomenclature that assists users in easily identifying the data sources in polynomials representing the provenance of distributed query results. 

To the best of our knowledge, our methods are the first to generate provenance polynomials for a wide range of SQL operations, including SPJUA, and correlated and uncorrelated nested queries without needing to change the structure of the original query results. It can seamlessly interact with different DBMSs and outperforms previous work in performance and scalability for many queries listed in the TPC-H benchmark. It is also the first solution that fully implements the aggregation method presented in~\cite{Yael2011}. For validation purposes, we implemented these methods in a software called DataProv. The main contributions are:

\begin{itemize}
    \item A novel definition of the rewriting rules for annotating SPJUA operations using only relational algebra operations that comply with the formalism proposed in~\cite{Green2007,Yael2011}. It is an advance over Perm and GProM, which do not present the annotations as provenance polynomials and do not provide information about aggregations. Unlike ProvSQL, DataProv is DBMS-independent and supports nested queries.

   \item A method to build and propagate the provenance of SPJUA operations and nested queries as provenance polynomials while preserving the structure of the original query results. 
   We also expand previous work by allowing users to choose which they want to see the complete information, including tuples that do not appear in the query result but are used in the computation process, or just the results of the original query with the corresponding provenance information.

   \item  An evaluation of Perm, GProM, ProvSQL, and our methods using the TPC-H benchmark. Unlike previous works, which focus on a limited number of queries and separately evaluate each solution, our experiments provide a more comprehensive assessment and a comparison of the expressiveness and performance of those frameworks and the approaches they follow.
\end{itemize}

The remainder of this paper is organised as follows. Section~\ref{sec:related_work} reviews some background on data provenance and related works. Section~\ref{sec:notations} presents our annotation rules and the nested query transformations. Section~\ref{sec:dataprov} describes the DataProv architecture and how to generate provenance polynomials. Section \ref{sec:experimental} presents and discusses experimental results. Section \ref{sec:conclusion} concludes the work and outlines future works.

\section{Background and Related work}
\label{sec:related_work}

Tracking data history, from its creation to its use, has long been a concern of information system developers and users. During the 1990s, the concept of data lineage emerged~\cite{Buneman2018} and, since then, significant progress has been made in this field~\cite{Sheikh2018,Cui2000}. Due to the ever-increasing importance of understanding the history and origin of data, a broader concept has emerged, called provenance~\cite{Buneman2018,Herschel2017}. In~\cite{Herschel2017}, provenance is categorised into four types: metadata provenance, information system provenance, workflow provenance, and data provenance. The latter focuses on providing information at the database and query level an is the focus of this paper.

\begin{table}
    \caption{Running example}
     \label{tab:temps}
     \vspace{-0.2cm}
    \begin{subtable}{\linewidth}
       \centering
       \caption{TE\_azores}
       \label{tab:tept}
        \begin{tabular}{|c|c|c|c}
    \cline{1-3}
      \textbf{timestamp} & \textbf{sn} & \textbf{duration} &\\
      \cline{1-3}
      08:00:00.120 & sn123 & 100 & t1\\
      \cline{1-3}
      09:15:32.165 & sn234 & 150 & t2\\
      \cline{1-3}
      12:40:55.180 & sn345 & 220 & t3\\
      \cline{1-3}
      22:32:10.220 & sn123 & 100 & t4\\
      \cline{1-3}
       \end{tabular}
    \end{subtable}
    \hfill
    \begin{subtable}{\linewidth}
        \centering
       \caption{Equipments}
        \label{tab:equip}
        \begin{tabular}{|c|c|c}
    \cline{1-2}
      \textbf{sn} & \textbf{model} & \\
      \cline{1-2}
      sn123 & ModelA & t5\\
      \cline{1-2}
      sn234 & ModelA & t6\\
      \cline{1-2}
      sn345 & ModelB & t7\\
      \cline{1-2}
       \end{tabular}
     \end{subtable}
     \\
     \begin{subtable}{\linewidth}
        \centering
       \caption{TE\_madeira}
       \label{tab:temad} 
       \vspace{-0.5cm} 
       \begin{tabular}{|c|c|c|c|c}
    \multicolumn{5}{c}{\textbf{}} \\
      \cline{1-4}
      \textbf{sn} & \textbf{model} & \textbf{num\_events} & \textbf{total\_duration} & \\
      \cline{1-4}
      sn202 & ModelB & 10 & 7600 & t8\\
      \cline{1-4}
      sn206 & ModelB & 5 & 9105 & t9\\
      \cline{1-4}
     sn440 & ModelA & 7 & 9750 & t10\\
      \cline{1-4}
       \end{tabular}
       
     \end{subtable}
     
\end{table}

Table~\ref{tab:temps} presents a running example representing data on traffic events per equipment of a telecom operator.
\textit{TE\_azores} and \textit{Equipments} represent traffic events in the Azores region. The former includes the timestamp, the serial number of the equipment (\textit{sn}) and the duration of a traffic event,  e.g., a phone call. The latter represents the serial number (\textit{sn}) and the \textit{model} of the network equipment. \textit{TE\_madeira} represents the number (\textit{num\_events}) and total duration (\textit{sum\_duration}) of the traffic events in the Madeira region aggregated by network equipment.
Each tuple has a provenance token ($t_i$), e.g., a universally unique identifier~\cite{Green2007,Buneman2001}.

\subsection{Data provenance}
\label{sec:SoA_formalisms}

While other forms of provenance focus on entities with low granularity, such as documents, files, and database tables, data provenance provides information about each data item, such as a value or a tuple, in a query result, including the source and operations performed to create it~\cite{Herschel2017,Cheney2007}. As such, data provenance can be very useful in clarifying query results~\cite{Senellart2019}, such as to explain why a data item is in a query result, how it was generated or the origin of a particular value. This enables objective evaluations of data quality~\cite{Herschel2017} or helps with query debugging~\cite{DemoDProv_debugging}, and is also relevant in auditing, reproducibility, debugging and other use cases~\cite{Herschel2017,Green2010,Senellart2017,Senellart2018}.

In~\cite{Green2009,Green2017}, tuple-based provenance is categorised as a hierarchy with provenance polynomials (\(N[X]\)) at the top. Provenance polynomials yield Boolean semiring provenance (\(B[X]\)) and trio provenance (\(Trio(X)\)). These, in turn, lead to why-provenance (\(Why(X)\)). Positive boolean expressions (\(PosBool(X)\)) and lineage semirings (\(Lin(X)\)) can be derived from the latter.

There also are other types of tuple-based provenance in the literature. An example is the why-not-provenance~\cite{Bidoit2015}, designed to explain why a tuple is not in the query result, and \(Sorp(X)\)~\cite{Daniel2014},  a type of provenance that can be derived from \(B[X]\) and is used to get the so-called minimal bag-witnesses~\cite{Green2017}.

\textbf{Provenance polynomials (\(N[X]\))} are at the top of the tuple-based provenance hierarchy and provide the most detailed information about the tuples in the query results and how they were computed. 
Provenance polynomials were initially introduced in~\cite{Green2007} based on the concept of provenance semirings. A provenance semiring is an algebraic structure \(( \mathbb{K}, +, \cdot,   0, 1) \) that includes a set of provenance tokens identified by $\mathbb{K}$, a commutative and associative operator $+$ with identity  \( 0 \), and commutative and associative operator $\cdot$ with identity  \( 1 \), which is distributive over $+$~\cite{Green2007,Senellart2017,Buneman2018}. In the semiring-based provenance model, the constants \(1\) and \(0\) represent the multiplicative and additive identities, respectively. The annotation  \( 1 \) means that a data item contributes neutrally to provenance computations—for example, joining with a tuple annotated with  \( 1 \) preserves the annotation of the other tuple. In contrast,  \( 0 \) indicates absence or non-contribution, effectively eliminating any provenance when combined. This formalism supports SPJU queries, where $\cdot$ represents conjunctions (such as \texttt{Join} and \texttt{Cartesian Product}) and $+$ represents alternatives (such as \texttt{Union}, \texttt{Projection}, and \texttt{Aggregation}).

For instance, consider Query~\ref{qy:exSum} and the instances shown in Table~\ref{tab:temps}:

\begin{equation}
    \begin{gathered}
     \myBigGamma_{model,SUM(duration) \rightarrow total}  (\myBigSigma_{TA.sn = E.sn} (TA \times E))
     \end{gathered}
    \label{qy:ex_poly}       
\end{equation}

\noindent
where $TA$ refers to the relation \texttt{TE\_azores}, and $E$ refers to \texttt{Equipments}. The result of the query is presented in Table~\ref{tab:ex_poly}, including the provenance polynomials. 
The provenance polynomial for ``ModelA'' is 
$(t_1 \cdot t_5) + (t_2 \cdot t_6) + (t_4 \cdot t_5)$, which indicates that three \texttt{Join} operations contributed to the result, between tuples $t_1$ and $t_5$, $t_2$ and $t_6$, and $t_4$ and $t_5$, respectively. The operator ($\cdot$) denotes the \texttt{Join} between two tuples, while ($+$) denotes alternative contributions to the aggregate value.

\begin{table}
    \centering    
    \caption{The result of Query~\ref{qy:ex_poly}}
   \label{tab:ex_poly}
    \begin{tabular}{|c|c|c|}
        \hline
        \textbf{sn} & \textbf{total} & \textbf{$\mathbb{N}[X]$} \\ 
        \hline
        ModelA & 350 & $(t_1 . t_5) + (t_2 . t_6) + (t_4 . t_5)$ \\
        \hline
        ModelB & 220 & $(t_3 . t_7)$ \\
        \hline
   \end{tabular}
\end{table}

As shown in Table~\ref{tab:ex_poly}, provenance polynomials do not capture aggregation information. To address this limitation, the formalism is extended in~\cite{Yael2011} to support the annotation of \texttt{Aggre-\allowbreak gation} functions such as \texttt{Count}, \texttt{Sum}, \texttt{Max}, and \texttt{Min}, by introducing a semimodule structure defined over a semiring and a monoid. The semimodule is constructed as a tensor product $K \otimes M$, where $K = \mathbb{N}[X]$ is the commutative semiring of provenance polynomials over a set of tokens X, and M is a commutative monoid representing the aggregation domain. Aggregation operators must be commutative and associative, as these properties are required for the result domain to form a commutative monoid. This ensures that the aggregation semantics can be embedded into the semimodule structure $K \otimes M$, and that query evaluation remains well-defined and consistent with provenance propagation. For instance, when modelling summation, one may use the monoid $M = (\mathbb{R}, +, 0)$, where $\mathbb{R}$ denotes the set of real numbers.

To capture the presence condition of each group in aggregation results, the framework employs a $\delta$-semiring, which extends \( K \) with a unary operator $\delta_K$. In the case of provenance polynomials \( \mathbb{N}[X] \), this operator maps the zero polynomial to \( 0 \) and any non-zero polynomial to \( 1 \), generalising duplicate elimination. This construction ensures compatibility with semiring homomorphisms, preserving a key correctness property of the model. The result of Query~\ref{qy:ex_poly} extended with~\cite{Yael2011} is present in Table~\ref{tab:ex_poly_yael}. Here, each value in the column total belongs to the semimodule $\mathbb{N}[X] \otimes M$, meaning it pairs a numeric aggregation with its symbolic provenance. The expression $(t_1 \cdot t_5) \otimes 100$, for example, indicates that the value 100 originates from the join of tuples $t_1$ and $t_5$. In the third column, the provenance polynomials now include the $\delta_{K}$ operator. These symbolic results preserve the full derivation history of each aggregate and can later be evaluated numerically via a semiring homomorphism (e.g., h: $\mathbb{N}[X] \to \mathbb{N}$).

\begin{table}
    \centering    
    \caption{The result of Query~\ref{qy:ex_poly} extended with~\cite{Yael2011}}
   \label{tab:ex_poly_yael}
    \begin{tabularx}{\linewidth}{|c|Y|Y|}
        \hline
        \textbf{sn} & \textbf{total} & \textbf{$\mathbb{N}[X]$} \\ 
        \hline
        ModelA & $(t_1 \cdot t_5) \otimes 100 +_{K \otimes SUM} (t_2 \cdot t_6) \otimes 150 +_{K \otimes SUM} (t_4 \cdot t_5) \otimes 100$ & $\delta_K((t_1 \cdot t_5) +_K (t_2 \cdot t_6) +_K (t_4 \cdot t_5))$ \\
        \hline
        ModelB & $(t_1 \cdot t_5) \otimes 220$ & $\delta_K((t_3 . t_7))$ \\
        \hline
   \end{tabularx}
\end{table}

To support nested aggregation queries, the semantics is extended to allow symbolic expressions of the form \( [x \, \alpha \, y] \), where \( x, y \in K \otimes M \), and \( \alpha \in \{=, \neq, <, \leq, >, \geq\} \). These expressions represent comparisons involving aggregate results, where \( x \) and \( y \) may be either annotated values or embedded constants. When a constant \( m \in M \) appears, it is embedded using \( \iota(m) = 1_K \otimes m \), enabling uniform treatment within the tensor product space. For instance, if  $y$  corresponds to  $m$ , the expression becomes  $[x \, \alpha \, 1_K \otimes m]$. Upon valuation of the provenance tokens, these symbolic expressions are resolved to \( 1_K \) if the comparison holds, and \( 0_K \) otherwise.

As an example, reconsider Query~\ref{qy:ex_poly}, now extended with a \texttt{Selection} condition to get only tuples with \texttt{total} greater than 150, i.e., $\sigma_{\text{total} > 150}$. The result for the ``ModelB'' is: 
\begin{align*}\delta_K((t_3 . t_7)) \textcolor[HTML]{0072B2}{\cdot_K [(t_1 \cdot t_5) \otimes 220 > 1_k \otimes 150]}\end{align*}

The symbolic comparison $[x > 1_K \otimes 150]$ is embedded in the provenance annotation (in \textcolor[HTML]{0072B2}{\textbf{blue}}) and evaluated alongside the presence condition $\delta_K$. When a semiring homomorphism $h : K \to K'$ is applied (e.g., assigning numeric values to provenance tokens), these symbolic comparisons can be resolved to either $1_K$ or $0_K$, depending on whether the predicate holds. This also captures the reason a tuple was included or excluded based on the filter.

When aggregating over values already annotated in $K \otimes M$, the relation must be lifted from a $K$-relation over $M$ to one over $K \otimes M$. This is achieved using the embedding $\iota : M \rightarrow K \otimes M$, where each $m \in M$ is mapped to $1_K \otimes m$. To pair provenance annotations with embedded values, the framework introduces the operator $*_{K \otimes M}$. Therefore, if we apply a \texttt{Sum} operation over the ``total'' column in the result of the \texttt{Selection} performed in Query~\ref{qy:ex_poly}, we obtain the following annotation in $\mathbb{N}[X]$: 

\begin{align*}
& A \textcolor[HTML]{0072B2}{\cdot_K [(t_1 \cdot t_5) \otimes 100 +_{K\otimes SUM} (t_2 \cdot t_6) \otimes 150 +_{K \otimes SUM} (t_4 \cdot t_5) \otimes 100} \\ 
& \textcolor[HTML]{0072B2}{> 1_k \otimes 150]}  \textcolor[HTML]{E69F00}{*_{K\otimes SUM} ((t_1 \cdot t_5) \otimes 100 +_{K \otimes SUM} (t_2 \cdot t_6) \otimes 150} \\
& \textcolor[HTML]{E69F00}{+_{K \otimes SUM} (t_4 \cdot t_5) \otimes 100)} +_{K \otimes SUM} B
\textcolor[HTML]{0072B2}{\cdot_K [(t_1 \cdot t_5) \otimes 220 > 1_k \otimes 150]} \\
& \textcolor[HTML]{E69F00}{*_{K\otimes SUM} (t_1 \cdot t_5) \otimes 220}
\end{align*}

\noindent
where, for simplicity, we denote by $A$ the value of the $\mathbb{N}[X]$ column in Table~\ref{tab:ex_poly_yael} corresponding to ``ModelA'', and by B the value corresponding to ``ModelB''. The text in \textcolor[HTML]{E69F00}{\textbf{orange}}, denotes the use of the operator $*_{K \otimes M}$, which lifts values from the monoid $M$ into the semimodule $K \otimes M $, to support aggregation over aggregated values.

\subsection{Related Work}
\label{sec:SoA_implementations}

Orchestra~\cite{Green2010} is an early solution for capturing provenance at the database level, with the goal of enabling the mapping of the schema of various sources for data sharing. It uses a specific language called ProQL, which does not handle general-purpose SQL queries, to express the provenance information. LogicBlox~\cite{Aref2015} is a solution designed for Datalog that also supports provenance annotations in the style of semirings for query debugging. 

In~\cite{Elhalawati2022}, the authors present a framework to calculate why-provenance using Datalog. In~\cite{Daniel2014}, the focus is on an extendend version of $PosBool(X)$ and on boolean provenance circuits.

Perm~\cite{Glavic2009} is a framework designed to build provenance annotations in PostgreSQL that handles a large subset of SQL, including SPJUA operations, nested queries and user-defined functions. User queries are rewritten into provenance-enabled queries using only SQL. It uses the  Perm-Influence Contribution Semantics (PI-CS) for why-provenance and the Copy Contribution Semantics (C-CS) for where-provenance. 
The same laboratory developed GProM~\cite{Arab2018}, a middleware that is not tied to any specific DBMS. GProM implements how-, why- and why-not-provenance. 
GProM can also export provenance annotations in W3C PROV for interoperability with other systems.

PUG~\cite{PUG_glavic} is an extension of GProM that implements a provenance model to represent the provenance of Datalog queries. The proposed provenance model allows for obtaining semiring provenance from positive queries.

One of the most recent approaches to capture provenance information is ProvSQL, which computes two types of provenance (how- and where-) and probabilities as an extension to PostgreSQL~\cite{Senellart2017,Senellart2018,Senellart2019}. The provenance annotations are appended to the query results as additional columns using PL/pgSQL (the procedural programming language of PostgreSQL), C, and C++. ProvSQL includes a query rewriting module that utilises provenance tokens as identifiers for gates in a provenance circuit, enabling the computation of provenance polynomials. Additionally, it provides user-defined functions that allow users to retrieve various types of provenance information. Currently, ProvSQL supports SPJUA and set operations, but it handles just some types of sub-queries and formulas. 

With the growing interest in data provenance, more solutions have appeared, such as SPARQLprov~\cite{Sparqlprov}, a middleware to compute how-provenance polynomials for monotonic and non-monotonic SPARQL queries and lineage annotations for queries with aggregates. The architecture is similar to GProM, and its theoretical foundations are close to the ones used by ProvSQL. Other works on data provenance in non-relational databases are Titian, a library that enables data lineage tracking in Apache Spark~\cite{ProvSpark}, and a method for witness generation for JSON schema~\cite{Scherzinger2019}. Additionally, the optimisation of provenance computing tasks remains a significant challenge. Although there is some work in this area~\cite{Niu2019}, further research is needed to find methods to decrease the overhead of computing provenance information. 

\subsection{Summary}

Several formalisms have been proposed to deal with different types of data provenance, with provenance polynomials being particularly promising, as it encompasses other types of provenance. However, the representation of provenance polynomials has not yet been fully achieved.
Perm and GProM can handle a wide range of SQL operations and generate witnesses of a query result, but cannot generate annotations as illustrated below. Also, as they are not designed to aggregate multiple witnesses for an output tuple, query results with provenance have to be disaggregated. Both solutions cannot deal with the provenance information for aggregations.

ProvSQL is capable of generating annotations based on provenance polynomials for both monotonic and non-monotonic queries. Nonetheless, it exhibits certain limitations: it does not support nested queries, it can only process \texttt{Having} clauses when they involve simple comparison predicates, and it does not provide support for \texttt{Order by} operations on aggregated columns.

To illustrate the differences between various solutions, consider the query in Listing~\ref{qy:exSum} and the results presented in Tables~\ref{tab:perm_gprom} and~\ref{tab:provsql}.

\begin{lstlisting}[mathescape=true, language=SQL, caption={An example of a query over the instances in Table~\ref{tab:temps}}, label={qy:exSum}]
SELECT sn, SUM(duration) as total
FROM te_azores
GROUP BY sn
\end{lstlisting}

\begin{table}
    \centering
    \caption{The result of Listing~\ref{qy:exSum} with Perm and GProM}
   \label{tab:perm_gprom}
   \vspace{-0.3cm}
    \begin{tabularx}{\linewidth}{|c|c|Y|Y|Y|}
      \hline
  \textbf{sn} & \textbf{total} & \textbf{prov\_te \_azores\_ timestamp} & \textbf{prov\_te \_azore\_sn} & \textbf{prov\_te \_azores \_duration} \\ 
    \hline
   sn345 & 220 & 12:40:55.180 & sn345 & 220 \\
     \hline
   sn123 & 200 & 08:00:00.120 & sn123 & 100 \\
     \hline
   sn123 & 200 & 22:32:10.220 & sn123 & 100 \\
     \hline
   sn234 & 150 & 09:15:32.165 & sn234 & 150 \\
     \hline
   \end{tabularx}
\end{table}

\begin{table}
    \centering    
\caption{The result of Listing~\ref{qy:exSum} with ProvSQL}
   \label{tab:provsql}
    \begin{tabularx}{\linewidth}{|c|c|Y|Y|}
        \hline
        \textbf{sn} & \textbf{total} & \textbf{how\_provenance} & \textbf{agg\_info} \\ 
        \hline
        sn123 & 200 & $\delta((t1 \oplus t4)) $ & $sum\{ (100 * t1) , (100 * t4) \}$ \\
        \hline
        sn234 & 150 & $\delta(t2)$ & $sum\{ (150 * t2) \}$ \\
        \hline
        sn345  & 220 & $\delta(t3)$ & $sum\{ (200 * t3) \}$ \\
        \hline
   \end{tabularx}
\end{table}

In both cases, the first two columns represent the query results, while the remaining columns contain provenance annotations. For Perm and GProM, the provenance consists of the values of the four tuples that contributed to the result (Table~\ref{tab:perm_gprom}). However, the structure of the original query result is altered, as it duplicates the entry for ``sn123'', which would otherwise appear only once. It is also important to note that GProM, unlike Perm, offers the capability to reduce the number of columns. Users must issue a command to select which column they wish to display as part of the provenance result.

In ProvSQL (Table~\ref{tab:provsql}), the provenance is built using the tuple identifiers and is represented as provenance polynomials. The aggregation information is represented as a separate column. This representation is more comprehensive, concise, and easier to interpret, aligning more closely with the formalism proposed in~\cite{Green2007,Green2017,Yael2011}.

ProvSQL also captures information about the aggregations performed, such as a \textit{SUM} in the example. In contrast, both Perm and GProM require prior knowledge of the query semantics to interpret how aggregate values are computed. Although ProvSQL does not fully adhere to the method proposed in~\cite{Yael2011}, introducing some variation in the representation of results, it was the first solution to offer a comprehensive approach to handling aggregations. However, it is unable to accurately capture the provenance of nested queries, which limits its applicability in more complex query scenarios. Furthermore, as it is implemented as an extension of PostgreSQL, it is restricted to that specific DBMS. All the approaches mentioned, require either additional query instrumentation or auxiliary data structures (e.g., extra tables or views) to compute and represent provenance information.
\section{Data provenance annotations}
\label{sec:notations}

Our method transforms SQL queries $\mathcal{Q}$ into annotated queries $\mathcal{Q}^+$ that use standard relational operations to build a provenance annotation for each tuple in a query result. It presents the provenance information as semirings~\cite{Green2007,Cheney2007,Senellart2018} and monoids~\cite{Yael2011}. The annotation rules allow the propagation of the provenance through the algebraic expressions in the query. The result of a query $R(a_1, \dots, a_n)$ is transformed into $R^+(a_1, \dots, a_n, prov)$, where \textit{prov} is a provenance annotation. Columns with aggregated values also have a specific algebraic annotation and are denoted as $a'_n$ and in these cases the corresponding relations are denoted $R'^+$.

We define the annotation rules for each relational algebra operator. These rules constitute the basis of our query rewriting method. We also present the rules used to transform nested queries, converting them into basic relational operations,  ensuring that there is no loss of provenance information. 
We also show how to apply these rules recursively to every operator in a tree, to build the provenance annotations step by step. In the remainder of this work, we consider the following:

\begin{itemize}
    \item \textbf{Token}. Each tuple $i$ in a relation $R$ is uniquely identified by a token $t_i$ (e.g., a Universally Unique Identifier - UUID).
    
    \item \textbf{Annotated relation}. $R^+$ is a base relation $R$ expanded with a column named $prov$ that has the structure of a semiring~\cite{Green2007}. The values of the aggregations are also annotated and have the structure of a monoid~\cite{Yael2011}.

    \item \textbf{Provenance annotations (\textit{prov})}. $R$ may be a base relation or an intermediate relation in the execution of a query. In the former case, $R^+.prov_i = R^+.t_i$, while in the latter, $R^+.prov_i$ is a composition of provenance tokens resulting from successive application of the transformation rules defined in this section. 
    Thus, $R^+.prov_i$ refers to the provenance annotation of the i-th tuple in $R^+$. 
    
    \item \textbf{Set and Bag semantics}. The \textit{set semantics} is the default, but some operations use the \textit{bag semantics}. To distinguish them, in the following, we denote a bag \texttt{Union} and a \texttt{Projection} with duplicates as $\myUnionB$ and $\myPiB$, respectively.

    \item \textbf{Manipulation of provenance annotations}. Two operators are used to combine provenance tokens: \textit{concat} and an aggregation function. The former concatenates two or more strings provided as parameters, e.g., $s = \text{concat}(s_1,$ $\dots, s_n)$. The latter performs string aggregation across multiple tuples and is denoted abstractly as \texttt{func}, representing functions such as \texttt{listagg} or \texttt{string\_agg}, depending on the specific DBMS~\cite{Michels2018}. We represent this operation as \( \upgamma_{\texttt{func}(s, p)}(R) \), where $s$ is the attribute to be aggregated (e.g., provenance tokens), and $p$ is the delimiter (e.g., $+$). For instance, given a relation $R^+$ with three annotated tuples, the result of the aggregation would be $t_i + t_j + t_k$.

\end{itemize}

\subsection{Annotations rules for SPJUA operators}
\label{sec:SPJUAops}

Next, we present the transformation rules for each SPJUA operator. The left-hand side represents a relational algebra operation and the right-hand the transformation that should be implemented to build the provenance annotations.

\paragraph{\textbf{Selection}}
This operator selects a subset of the tuples of a relation that satisfy a given condition. As it is assumed that the base relations are initially transformed into annotated relations, there is no need for additional annotations.

\paragraph{\textbf{Projection}}
By default, a \texttt{Projection} of $u = (a_1, \ldots, a_n)$ removes duplicate entries. However, to preserve provenance information, the annotations must capture all input tuples that contribute to each projected result. This requires using aggregation functions to combine provenance tokens corresponding to identical projected values. As a result, the projected attributes $(a_1, \ldots, a_n)$ must appear in the \texttt{Group By} clause. Consequently, the set \texttt{Projection} is implemented as a \texttt{Group By} operation. This transformation is further elaborated in the Aggregations section.
When bag semantics is used, all tuples are included in the result, including duplicates. Therefore, there is no need to change the provenance annotations, since each tuple already has its own provenance value. However, it is necessary to add the \textit{prov} column from $R^+$ (see~(\ref{qy:proj2})), which would otherwise be lost due to \texttt{Projection}.

\begin{equation}
         \myBigPiB_{u}(R^+) \stackrel{+}\Rightarrow \myBigPiB_{u\, , \, prov \,}(R^+)
        \label{qy:proj2}       
\end{equation}

\paragraph{\textbf{Join}}

The annotation of \texttt{Join} (or \texttt{Cartesian Product}) operations is a conjunction of the provenance annotations of the tuples from the participating 
relations using the operator ``$\cdot$''. 

\begin{equation}
    \begin{gathered}
        R^+ \, \myBigJoin \, _{condition} \, V^+ \stackrel{+}\Rightarrow \\ \, \myBigPi_{u \, , \, concat \big(\, R^+.prov, '\cdot', V^+.prov \big) \rightarrow prov} \Big( R^+ \, \myBigJoin \, _{condition} \, V^+ \Big) 
    \end{gathered}
    \label{qy:join}
\end{equation}

For example, the output of a natural \texttt{Join} with a \texttt{Selection}
$\mySmallSigma_{model = 'ModelB'}
(TE\_azores \,\, \myBigJoin \,\, Equipments)$ 
is
\textit{\{(12:40: 55.180, sn345, 220, sn345, ModelB, $(t3 \cdot t7)$)\}}.

\paragraph{\textbf{Union}} 
As the \textit{Projection} operator, the \textit{Union} can be interpreted with either set or bag semantics. In the case of bag semantics, all tuples from all relations are included in the result without requiring additional annotations. When the \textit{Union} operates as a set operator, 
it is first converted into a bag \textit{Union} and then enclosed within a \textit{Group By} clause, which eliminates duplicates and enables the aggregation of the provenance of duplicated tuples using \textit{func}. To avoid ambiguities and ensure consistent referencing of the attributes in the new bag \textit{Union} relation, a rename operation is also included as shown in~(\ref{qy:union1}).

\begin{equation}
    \begin{gathered}
        R^+ \, \myBigCup \, V^+ \stackrel{+}\Rightarrow 
        \\ \myBigGamma_{U^+.u, \, func\big(U^+.prov, '+'\big) \rightarrow prov} \, \myBigRho(U^+) \, \Big( R^+ \, \myBigUnionB \, V^+ \Big)
    \end{gathered}
    \label{qy:union1}
\end{equation}

For example, if $R = \myMediumPi_{model} (Equipments)$ and $V = \myMediumPi_{model} \\(TE\_madeira)$, then the output of 
$R \, \myBigCup \, V$ 
is
\textit{\{(ModelA, $t5 + t6 + t10$), (ModelB, $t7 + t8 + t9$ )\}}.

\paragraph{\textbf{Aggregation}} Aggregation functions such as \texttt{Sum}, \texttt{Avg}, \texttt{Count}, \texttt{Min} and  \texttt{Max} are collectively referred to as \texttt{AGG} for the sake of generalization. 
When the aggregation has not a  \texttt{Group By} clause, the transformation is as shown in ~\ref{qy:agg1}.

\begin{equation}
    \begin{gathered}
        \myBigGamma_{AGG(a_n) \rightarrow a_n'} (R^+) \stackrel{+}\Rightarrow  \\ \myBigGamma_{func(concat(prov,'\otimes', a_n), '+AGG') \rightarrow a_n'} (R^+)    
    \end{gathered}
    \label{qy:agg1}
\end{equation}

For instance, $\myGamma_{SUM(num\_events) \rightarrow total} (TE\_madeira)$ gives \textit{\{($t8 \otimes 10 +_{sum} \, t9 \otimes 5 +_{sum} \, t10 \otimes 7$)\}}, such that $\otimes$ stands for scalar multiplication. The notation $+_{sum}$ is used in place of the $+_{k \otimes sum}$ proposed in~\cite{Yael2011}, to simplify the expression.
Note that, in this case, the aggregate value ($22$ in the example) is replaced by a monoid that shows how the value was computed.

When there is a \texttt{Group by}, the aggregations are annotated with  $\delta$  representing a commutative $\delta$-semiring~\cite{Yael2011}. The transformation is shown in~(\ref{qy:gp1}).

\begin{equation}
 \myBigGamma_{u}(R^+)  \stackrel{+}\Rightarrow \, \myBigGamma_{u, \, concat\big( \,'\delta(', func(prov, '+'), ')'\big) \rightarrow prov} \, (R^+)
    \label{qy:gp1}
\end{equation}

Therefore, the result of $\myGamma_{sn}(TE\_azores)$ is \textit{\{(sn123, $\delta(t1 + t4)$), (sn234, $\delta(t2)$), (sn345, $\delta(t3)$)\}}, where $+$ denotes an alternative, i.e., \textit{sn123} can be derived from either $t_1$ or $t_4$.
This is also the transformation that set \texttt{Projections}, thus:

\begin{equation}
	\myBigPi_{u}(R^+) \stackrel{+}\Rightarrow \Big( \myBigGamma_{u}(R^+) \Big)^+ 
	\label{qy:aggProj}
\end{equation}

The transformation for a \texttt{Group By} with an aggregation function 
is a combination of (\ref{qy:agg1}) and (\ref{qy:gp1}), and is represented in (\ref{qy:aggProjGB}), where $a'_n$ has the structure of a monoid representing an aggregation as defined in~(\ref{qy:agg1}) and
$prov$ has the structure of the $\delta-$\textit{semiring} built in~(\ref{qy:gp1}).

\begin{equation}
	\myBigGamma_{u, \, AGG(a_n)}(R^+) \stackrel{+}\Rightarrow \myBigGamma_{u, \,\, a'_n, \,\, prov'}(R^+)
	\label{qy:aggProjGB}
\end{equation}

This is the case of query in Listing~\ref{qy:exSum}, and the result following the rules presented in this section is present in Table~\ref{tab:dataprovex}.

\begin{table}
    \centering    
\caption{The result of Listing~\ref{qy:exSum} with the rules of Section~\ref{sec:SPJUAops}}
   \label{tab:dataprovex}
    \begin{tabularx}{\linewidth}{|c|Y|Y|}
        \hline
        \textbf{sn} & \textbf{total} & \textbf{prov} \\ 
        \hline
        sn123 & $t1 \otimes 100 +_{sum} t4 \otimes 100$ & $\delta(t1 \oplus t4) $ \\
        \hline
        sn234 & $t2 \otimes 150$ & $\delta(t2)$ \\
        \hline
        sn345  & $t3 \otimes 220$ & $\delta(t3)$ \\
        \hline
   \end{tabularx}
\end{table}

\subsection{Extended semantics} 
\label{sec:ExtendedSemantics}
In earlier cases, it is assumed that the \textit{prov} column follows the structure of a $K$-semiring~\cite{Green2007}, while the remaining columns contain values without any predefined algebraic structure. However, once aggregation operations are annotated, this assumption no longer holds. Aggregated columns adopt the structure of a monoid, and the \textit{prov} column must be interpreted as a $\delta$-semiring~\cite{Yael2011}, which supports symbolic expressions and conditional annotations over aggregated values.

To generalise the framework, it becomes necessary to account for comparisons involving aggregate values and for nested aggregations. Let $R'^+$ denote the result of a query involving aggregation or selection over aggregated values. Then, the column $R^+.\textit{prov}$ may already include complex annotations involving $\delta$-semiring terms, symbolic comparisons, and lifted values in $K \otimes M$. These annotations must be preserved and correctly propagated through any subsequent operations.

\paragraph{\textbf{Selection}} 
In this scenario, the comparison is over an attribute resulting from an aggregation.
As proposed in~\cite{Yael2011}, selections on aggregated values are denoted by symbolic expressions represented by square brackets and all tuples are included, even when the selection condition does not hold. The transformation to add the annotation of the selection operations to the provenance information is: 

\begin{equation}
    \begin{gathered}
    \myBigSigma_{a'_n \, \alpha \, m}(R'^+) \stackrel{+}\Rightarrow
    \myBigPiB_{u, \, a'_n, \, concat\big(R'^+.prov, '\cdot[', a'_n, ' \alpha \, 1 \, \otimes ', m, ']'\big) \rightarrow prov} \,(R'^+)        
    \end{gathered}
    \label{qy:filter}
\end{equation}

\noindent 
where $a'_n$ represents the aggregated values, $\alpha$ is a relational operator, and $m$ is a value. 
In the expression on the right, $u$ denotes the attributes of $R'^+$ except $a'_n$ and $prov$, which are explicitly represented in the formula, to make clear that both input and output include a monoid $(a'_n)$ and a $\delta-$\textit{semiring} (\textit{prov}).

For instance, if we make a \texttt{Selection} with a condition on the query in Listing~\ref{qy:exSum} to select only tuples where the total is less than or equal to $200$, the query yields the result shown in Table~\ref{tab:filterex}.
The right-hand side is $1 \otimes 200$ for each case.
Section~\ref{sec:dataprov} explains how to calculate the logical value of each of the conditions. 

\begin{table}
    \centering
    \caption{The result of \texttt{Selection} over Listing~\ref{qy:exFilter3} with a filter by column ``total''}
   \label{tab:filterex}
    \begin{tabularx}{\linewidth}{|X|X|X|}
      \hline
  \textbf{sn} & \textbf{total} & \textbf{prov} \\ 
    \hline
	sn123 & $t1 \otimes 100 +_{sum} t4 \otimes 100$ & $\delta(t1 + t4).[t1 \otimes 100 +_{sum} t4 \otimes 100 <= 1 \otimes 200 ]$ \\
    \hline
    sn234 & $t2 \otimes 150$ & $ \delta(t2) . [t2 \otimes 150 <= 1 \otimes 200]$ \\
    \hline
    sn345 & $t3 \otimes 220$ & $\delta(t3) . [t3 \otimes 220 <= 1 \otimes 200]$ \\
     \hline
   \end{tabularx}
\end{table}

\paragraph{\textbf{Join}} 

As this is a Cartesian Product with a selection, these are handled similarly to the previous case (see \ref{qy:joinna}).

\begin{equation}
    \begin{gathered}
       R'^+ \myBigJoin_{a'_{n} \, \alpha \, b'_{n}} V'^+ \stackrel{+}\Rightarrow 
        \myBigPiB_{u, \, a'_n, \, v, \, b'_n,} \\ _{concat\big(R'^+.prov, '\cdot', V'^+.prov, '\cdot[', a_n', ' \alpha ', b_n',']'\big) \rightarrow prov} \big(R'^+ \times V'^+\big)
    \end{gathered}
    \label{qy:joinna}
\end{equation}

Let  $V'^+$ be a relation that includes \texttt{Aggregations}, $b'_n$ be an attribute that has been aggregated, and $v$ represent the attributes of $V'$ apart from $b'_n$. In this scenario, the symbolic expression consists of the comparison $a'_n \, \alpha \, b'_n$, such that $b'_n$ becomes a symbolic semimodule expression that includes symbolic comparisons instead of a comparison with a constant. However, if $b'_n$ were not an aggregated attribute, the representation of the right-hand side is a constant, as in a \texttt{Selection}. 

Note that these transformations only apply when the comparison in \texttt{Selection} or \texttt{Joins} are on aggregated values. In other cases the transformations listed in Section~\ref{sec:notations} apply.

\paragraph{\textbf{Projection and union}} 
For projections and unions there are no specific transformations in this scenario and therefore the rules defined in Section~\ref{sec:SPJUAops} apply.

\paragraph{\textbf{Aggregation}}
When an \texttt{Aggregation} is applied to a value that is already the result of a previous aggregation (i.e., annotated in $K \otimes M$), it is necessary to (i) update the monoid value represented by $a'_n$, and (ii) recompute the provenance annotations in the \textit{prov} column. This recomputation is done using the operator $*_{K \otimes M}$, which we abbreviate as $*_{\text{agg}}$, as shown in (\ref{qy:aggna}).

\begin{equation}
\begin{gathered}
\myBigGamma_{\text{AGG}(a'n)}(R^+) \stackrel{+}\Rightarrow \myBigGamma_{\text{func}(R'^+.a'_n, '+\text{agg}')} \\
_{\text{func}(\text{concat}(R'^+.prov, '*\text{agg}', R'^+.a'_n), '+') \rightarrow \text{prov}}(R'^+)
\end{gathered}
\label{qy:aggna}
\end{equation}

\paragraph{\textbf{Group By}}
When a \texttt{Group By} is applied to a relation that already contains aggregated values, and the grouping attributes $u$ do not include the aggregated column $a'_n$, it becomes necessary to propagate the existing aggregated values into the output. This is done by pairing them with their corresponding provenance annotations using $*_{\text{agg}}$, as illustrated in (\ref{qy:aggna2}).

\begin{equation}
\begin{gathered}
\myBigGamma_{u}(R^+) \stackrel{+}\Rightarrow \\
\myBigGamma_{u}, \
_{\text{func}(\text{concat}(R'^+.prov, '*\text{agg}', R'^+.a'_n), '+') \rightarrow \text{prov}}(R'^+)
\end{gathered}
\label{qy:aggna2}
\end{equation}

\subsection{Nested  queries}
\label{sec:nested}
 
When nested queries—also referred to as subqueries or inner queries—appear in the \texttt{Where} clause, their provenance cannot be captured, as they are not accessible from the outer query. As a result, it is not possible to construct the complete provenance polynomial as described in~\cite{Green2007}, nor to apply the extended formalism for aggregations proposed in~\cite{Yael2011}. Consequently, it becomes necessary to apply unnesting techniques that transform the nested subquery into an equivalent unnested query, enabling its integration into the outer query while preserving the original query semantics. In this way, it becomes possible to propagate provenance polynomials throughout the query and to apply the extended semiring–semimodule formalism to support aggregation annotations.

Nested queries can be correlated and uncorrelated. Uncorrelated queries are independent of the outer query and do not use any columns from it. In contrast, correlated queries are dependent and reference one or more outer query columns.

In the literature, various unnesting techniques have been proposed. In~\cite{Ganski1987}, the authors present methods that transform predicates such as \texttt{(NOT) Exists}, \texttt{(Not) In}, and \texttt{Any} or \texttt{All} into equivalent forms using scalar or set-containment operators such as \texttt{Count}, \texttt{Min}, or \texttt{Max}, without moving the nested query to the \texttt{From} clause. Thus, it would not possible to fully capture provenance polynomials from these transformations.

In~\cite{Neumann2015,Neumann2025}, the authors use left \texttt{SemiJoins} and \texttt{AntiJoins} to relocate correlated subqueries from the \texttt{Where} clause to the \texttt{From} clause, thereby eliminating dependent joins.

In~\cite{GlavicNested2009}, the authors introduce query rewriting strategies that transform queries featuring sublinks, including both correlated and nested subqueries, into equivalent forms through the replacement of sublinks with join operations. For uncorrelated sublinks, the \texttt{Left} strategy employs left outer joins to integrate the rewritten sublink queries, while additional strategies cater to correlated and more complex scenarios. This join-based approach is utilised in the Perm. These strategies cannot be directly applied in our method due to differences in the representation of provenance results.

We define the general structure of a nested query as $R^+ \, \Theta \, V^+$, such that $R^+$ stands for the outer query and $V^+$ the inner query, $\Theta \in \{ \alpha, \, \alpha \, Any, (Not )\, In, \, \alpha \, All, \, (Not) \, Exists \}$.

\paragraph{\textbf{Exists}} The \texttt{EXISTS} operator returns true if the subquery yields at least one row, stopping evaluation after the first match. In provenance polynomials, however, all tuples that could satisfy the condition must still be included.

When the correlation conditions involve only the equality operator (=), the \texttt{Exists} predicate can be effectively transformed into a \texttt{SemiJoin} between the outer relation and the inner relation, with a \texttt{Group By} applied to the attributes of the inner relation referenced in the correlated clauses to avoid duplicate contributions. This transformation is not directly applicable to inequalities, as these produce a range of matching values, which can lead to incorrect tuple replications in the result. To ensure a transformation that is valid for any type of correlation condition, the following generic approach is applied:

\begin{equation}
    \begin{gathered}
         \myBigSigma_{EXISTS (\myBigSigma_{condition} (V^+) )}(R^+) \rightarrow R^+ \, \myBigLeftSemijoin_{condition''} \myBigRho (V'^+) \\ (\myBigGamma_{nestedT.u} (V^+ \, \myBigRightSemijoin_{condition'} \, \myBigRho(nestedT) \, (\myBigPi_{u} R^+)))
    \end{gathered}
    \label{qy:exists}
\end{equation}

\noindent
such that, $\myBigSigma_{condition} (V^+)$ denotes a nested relation from the \texttt{Exists} predicate, where $u = (a_1 \dots a_n)$ and $v = (b_1 \dots b_n)$ denote the attributes defining the correlation between $R^+$ and $V^+$ in \textit{condition}. Inside of $V'^+$, the outer relation $R^+$ is projected on the attributes $u$ to form the intermediate relation ``\textit{nestedT}''. This projection ensures that each combination of correlated values is considered only once, avoiding unnecessary duplicates. A \texttt{right SemiJoin} is then performed between $V^+$ and ``\textit{nestedT}'' using \textit{condition}', which is identical to the original \textit{condition} between $u$ and $v$ attributes. The result of this \texttt{SemiJoin} is grouped by the attributes $u$ from ``\textit{nestedT}'' to consolidate all matching inner tuples into a single representative per correlation key, which reproduces the boolean semantics of \texttt{Exists} and enables provenance aggregation. Finally, a \texttt{left SemiJoin} is applied between $R^+$ and $V'^+$, where \textit{condition}'' uses the same correlation attributes as \textit{condition} but restricted to equality predicates.
This transformation is inspired by the \texttt{SemiJoin} decorrelation approach of~\cite{Neumann2015,Neumann2025} and the provenance-preserving \texttt{Join} strategies of~\cite{GlavicNested2009}, and is adapted for application within our method.

The resulting transformation is logically equivalent to the original query. However, the semantics of the query have changed with the inclusion of a \texttt{right SemiJoin} to create $V'^+$ that did not exist in the original query, and this would be reported in provenance annotations built using the rules listed in Section~\ref{sec:SPJUAops}. To ensure that the provenance annotations remain consistent with the original query, the relation is renamed to ``\textit{nestedT}'' and this exception is handled programatically, as explained in Section~\ref{sec:dataprov}.

\paragraph{\textbf{Any}} This operator can be correlated or uncorrelated. There are several alternatives to unnest queries with the \textit{Any} operator~\cite{Ganski1987} but they are not necessarily semantically equivalent and generate the same provenance information. In this work, we transform \textit{Any} into \textit{Exists} as follows:

\begin{equation}
     \myBigSigma_{a \, \alpha \, ANY \,(\myMediumPi_{b}(V^+))}(R^+)  = \myBigSigma_{EXISTS \, (\mySmallSigma_{condition} (V^+) )}(R^+)
 \label{qy:any1}
\end{equation}

\noindent
such that, \textit{condition} is equal to $R^+.a \, \alpha \, V^+.b$. If the operator \textit{Any} is correlated, then the join conditions are also incorporated into \textit{condition}.

\paragraph{\textbf{In}} The operator \texttt{In} may be correlated or uncorrelated. In the uncorrelated case, DataProv transforms the \textit{In} into a \texttt{SemiJoin} as proposed in~\cite{Neumann2015} as follows:

\begin{equation}
    \myBigSigma_{u \,\, IN \,(\myMediumPi_{v}(V^+))}(R^+)  \rightarrow R^+ \, \myBigLeftSemijoin_{ condition } \,\, \myBigRho (V') \big(\myBigGamma_v \, (V^+) \big)	
	\label{qy:inUN}
\end{equation}

\noindent
such that $u = (a_1 \dots a_n)$ and $v = (b_1 \dots b_n)$ denote attributes of $R^+$ and $V^+$, respectively. These attributes are used to define the join condition in the transformed query. To ensure that repetitions do not influence the result, a \texttt{Group By} clause is incorporated over the attributes of \( u \) within the relation \( V^+ \).

When the query is correlated the operator is transformed into an \textit{Exists}~\ref{qy:inCO}. 

\begin{equation}
 \begin{gathered}  
     \myBigSigma_{u \, IN \,(\mySmallSigma_{condition} \myMediumPi_{v}(V^+))}(R^+)  = \myBigSigma_{EXISTS } { \, (\mySmallSigma_{condition'} (V^+) )}(R^+)
     \end{gathered}
 \label{qy:inCO}
\end{equation}

\noindent
such that, \textit{condition'}~ incorporates an equality of $u$ and $v$ and the join condition between the outer and the nested queries.

\paragraph{\textbf{All}} This operator implies that the comparison is true for all tuples in the nested relation. Thus, the how-provenance information must aggregate them all. 
In~\cite{Ganski1987} it is shown that \texttt{All} operator can be transformed into \texttt{Max} or \texttt{Min} and joined with the outer query:

\begin{equation}
    \begin{gathered}    
    \myBigSigma_{a \,\, \alpha \, ALL \,(\myMediumPi_{b}(V^+))}(R^+) = \myBigSigma_{a \,\, \alpha \, (\myBigGamma_{AGG(b)}(\myMediumPi_{b}(V^+))}(R^+) \rightarrow
    \\ R^+ \myBigLeftSemijoin_{condition'} \,\myBigRho (V')( \myBigGamma_                  {AGG(b) \rightarrow b}(\myMediumPi_{b}(V^+)))
    \end{gathered}
    \label{qy:all}
\end{equation}

\noindent
such that, \textit{condition'}~ is $a \, \alpha \, b$.

\paragraph{\textbf{Comparison operator ($\alpha$)}} The nested queries with comparison operators follow almost always a straightforward transformation using the \texttt{SemiJoin} (\ref{qy:mathop}), where $condition' = {a \, \alpha \, b}$.

\begin{equation}
    \begin{gathered}
         \myBigSigma_{a \,\, \myBigAlpha \,\,(\myMediumPi_{b}(V^+))}(R^+) \rightarrow R^+ \,\myBigLeftSemijoin{condition'} \, (\myBigPi_{b} (V^+))
    \end{gathered}
    \label{qy:mathop}
\end{equation}

However, there is an exception when the nested query has a column with a \texttt{Count} or a \texttt{Max}, referred to as the ``Count Bug'' in~\cite{Ganski1987}. In this case, the transformation is similar to the approach followed in the \texttt{Exists} operator.

Since our method is based on a positive semiring algebra, it does not transform \texttt{Not In} and \texttt{Not Exists} constructs. Capturing the provenance of such non-monotone queries requires extending the provenance polynomial semiring $\mathbb{N}[X]$ with a subtraction operator, such as the monus ($\ominus$), as proposed in~\cite{Senellart2019}.

\section{The DataProv Middleware}
\label{sec:dataprov} 


We implemented our annotation strategy and the nested rules in a middleware 
called DataProv. This solution is independent of the DBMS. The source 
code for DataProv is available online. This section presents the architecture 
of DataProv, its key components, and the annotation propagation method.

\subsection{Architecture}

\begin{figure}
	\centering
	\includegraphics[width=\linewidth]{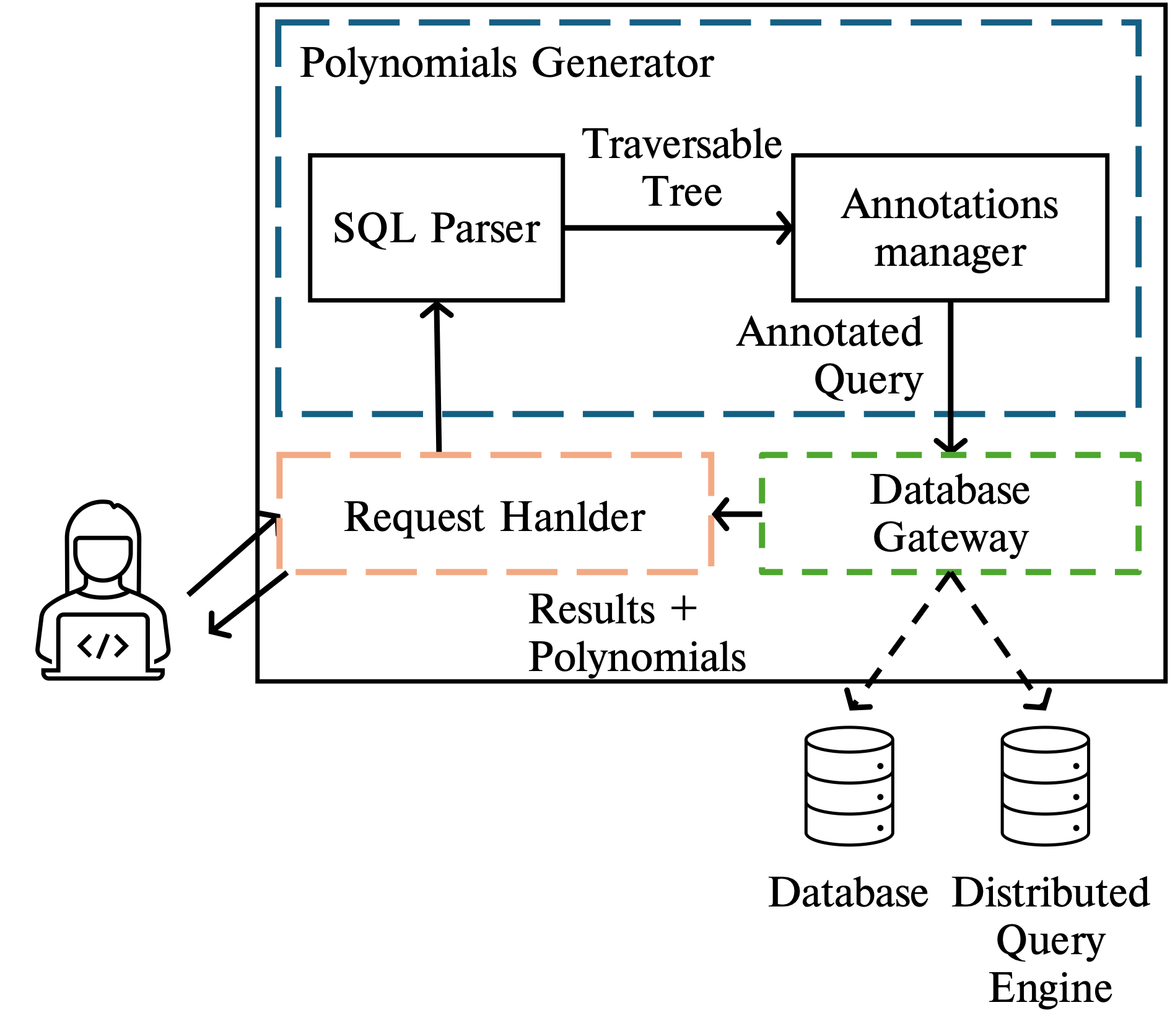}
	\caption{The DataPROV architecture}
    \label{fig:framework}
\end{figure}

Figure~\ref{fig:framework} presents the the architecture of DataProv. The \textit{Request Hanlder} receives a query and the information to connect to a database. The \textit{Polynomials generator} parses and modifies the queries with annotations before transmitting them to the Database Gateway. The \textit{Database Gateway} sends the annotated queries to the underlying DBMS or a Distributed Query Engine using Wrappers. The query execution result sent to the user includes the query execution resultset and the corresponding provenance information.

The \textbf{Polynomials generator} is implemented in JAVA and has two main components: an SQL parser and the \textit{Annotations Manager}. The SQL parser used in this work is an open-source library called JSQLParser\footnote{The JSQLParser website \url{https://jsqlparser.github.io/JSqlParser/index.html}.}, which generates a traversable tree of SQL objects that can be accessed and modified as long as the query is valid. The \textit{Annotations Manager} uses a recursive bottom-up approach to iterate on the parsed tree and add annotations, following the rules described in Sections~\ref{sec:SPJUAops} and \ref{sec:ExtendedSemantics}, from the lowest level operator to the top of the tree, enabling the propagation of provenance polynomials. This module also implements the unnesting rules described in Section~\ref{sec:nested}. Then, a new (annotated and transformed) SQL command is defined based on the modified tree and submitted for execution. It is important to note that, in distributed environments, provenance tokens follow the format with more information - \textit{db-Name\#schema\#table:$t_i$}, providing the user with complete information about the token’s origin.

The \textbf{Database Gateway} connects to DBMS and distributed query engines. Developed in JAVA, it uses Java Database Connectivity (JDBC) drivers to establish connections. Currently, the solution supports PostgreSQL, Oracle, and Trino, but it can be expanded to support other DBMSs and distributed query engines that accept the standard SQL. Trino is used to access multiple databases in distributed environments, including NoSQL databases.

\subsection{Polynomials generator}
\label{sec:how_prov_gen}

\begin{figure}
	\centering
	\includegraphics[width=\linewidth]{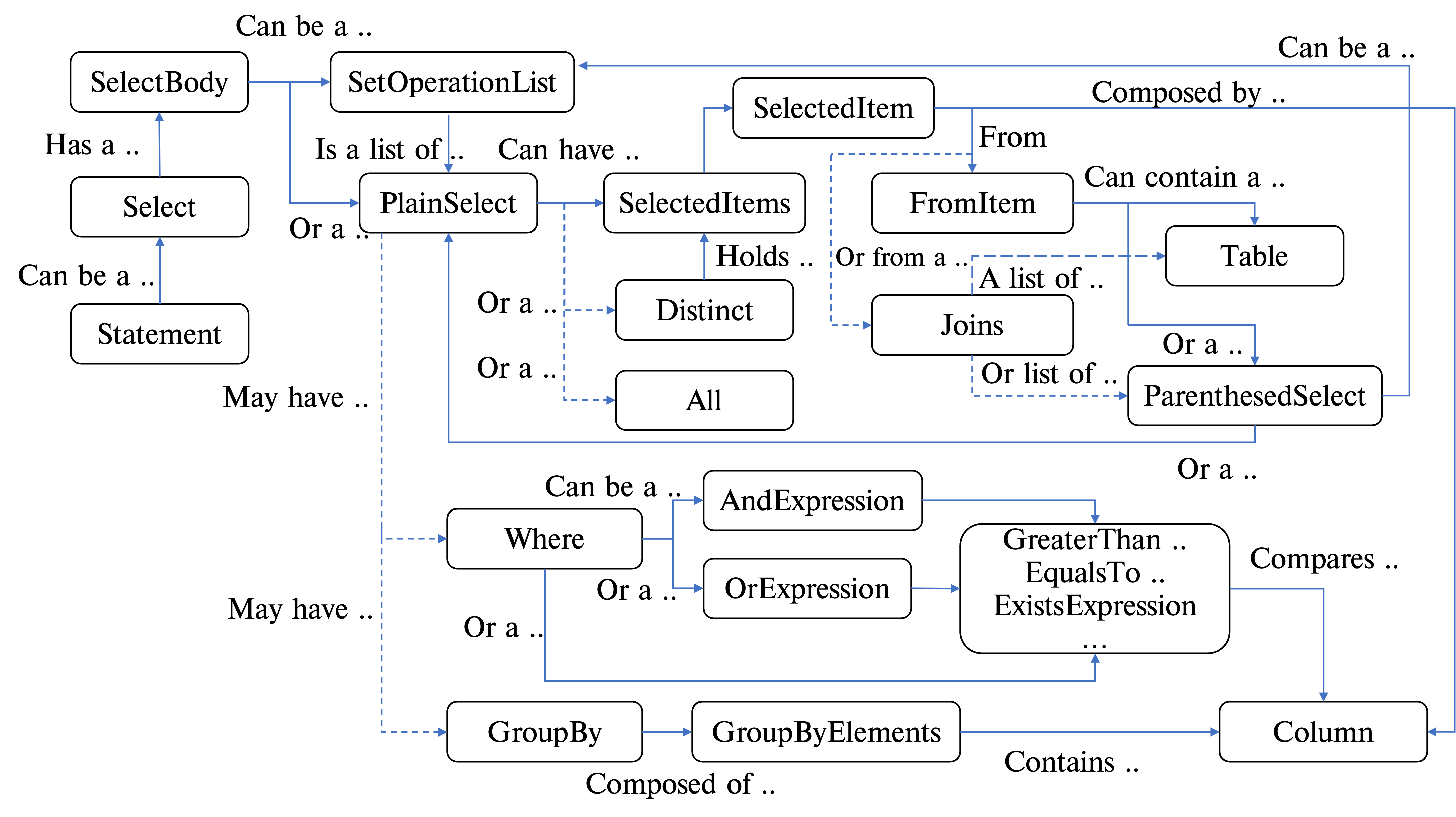}
	\caption{Main classes used in the traversable tree}
    \label{fig:JSQLParserobjects}
\end{figure}

Figure~\ref{fig:JSQLParserobjects} presents a visual representation of the JSQLParser objects and the syntax handled in DataProv.

The main class is ``Statement'', and a statement may be a ``Select'', which has a ``SelectBody''. A ``SelectBody'' can be a ``PlainSelect'' or a ``SetOperationList''. The former has items (``SelectedItens'') from ``Table'' or ``Joins'' and may also have ``Distinct'' or ``All''. The latter is a list of ``PlainSelect'' and can represent operations such as \texttt{Union}, \texttt{Union All} , \texttt{Intersect} and \texttt{Except}. A ``PlainSelect'' may also have a ``Where'' with logical expressions and subqueries and a ``GroupBy'' element. A ``Joins'' is a list that may have ``Table'' and ``ParenthesedSelect''. The latter, in turn, can be another ``SetOperationList'' or ``PlainSelect''. A ``SetOperationList'' may have another list of ``SetOperationList'' or ``PlainSelect''.

Algorithm~\ref{alg:addAnno} outlines the main steps for interpreting and transforming user queries into queries enabled to generate the provenance annotations and the order in which the rules listed in Section~\ref{sec:notations} are applied so that provenance annotations propagate correctly. This is the main function of the Annotations Manager 

\begin{algorithm}[ht]
\caption{The main function of \textit{Annotations Manager}}\label{alg:addAnno}
\footnotesize
\begin{algorithmic}[1]
\Function{addAnnotations}{$select$}
    \State $select \gets$ \Call{NestedTransform}{$select$}
    \If{$select$ instance of $SetOperationList$}
        \If{$select$ is (UNION or UNION ALL)}
            \For{each $object$ in $SetOperationList$}
                \State $object \gets$ addAnnotations($object$)
            \EndFor
            \State $select \gets$ \Call {UnionF} {$select$}
        \EndIf
    \Else 
        \If{$select.getJoins()$ is not NULL}
            \State $select \gets$ \Call {JoinF} {$select$}
        \Else
            \State $select \gets$ \Call {ProjF} {$select$}
        \EndIf

        \State $select \gets$ \Call {NestedAgg} {$select$}

        \If{$select.getGroupBy()$ is not NULL}
                \State $select \gets$ \Call {AggregationRules} {$select$}
        \EndIf
    \EndIf
    \State \Return $select$
\EndFunction
\end{algorithmic}
\end{algorithm}

The function ``AddAnnotations'' receives a query validated by the Parser and checks if it is a ``SetOperationList'', meaning a \texttt{Union} or \texttt{UnionAll} operation. If so, it calls the ``UnionF'' function to iterate through the list and applies the transformation Rule~(\ref{qy:union1}) Section~\ref{sec:SPJUAops}. Given that the list contains queries, it recursively calls the ``AddAnnotations'' function to implement the relevant transformations on the queries within the list. If the input is  a ``PlainSelect'', the algorithm calls the ``nestedTransform'' function. This function examines the presence of nested queries and applies the transformations detailed in Section~\ref{sec:nested}. Next, it checks if it is a \texttt{Join} and invokes the function named ``JoinF'' (Rule~\ref{qy:join}). The ``JoinF'' also leads to the exceptions mentioned in Section~\ref{sec:nested}. If it is not a \texttt{Join}, the function calls the ``ProjF'' to handle \texttt{Projections} (Rule~\ref{qy:proj2} - bag semantics) and \texttt{Selections}. Finally, it checks the presence of the \texttt{Group By} clause to implement the aggregation rules (Rules~\ref{qy:agg1}, ~\ref{qy:gp1} and~\ref{qy:aggProjGB}) and look for the keyword \texttt{Distinct}  to apply the Rule~\ref{qy:aggProj}, the \texttt{Projection} set semantics.

The subqueries are processed following a bottom-up recursive approach. The functions \textit{UnionF}, \textit{JoinF}, and \textit{ProjSelectF} also call the function \textit{addAnnotations} recursively.

\subsection{Examples}

\paragraph{Example 1.}
Consider the query in the Listing~\ref{qy:exFilter} with  an inner query that performs a projection and an aggregation, and an outer query that performs a selection on the aggregated values.

\begin{lstlisting}[language=SQL, caption={An example of a query with a \texttt{Selection} over the query in Listing~\ref{qy:exSum}}, label={qy:exFilter}]
SELECT *
FROM(SELECT sn, SUM(duration) total
        FROM te_azores
        GROUP BY sn) C0 WHERE C0.total > 200
\end{lstlisting}

The subquery in the \texttt{FROM} clause is processed first. According to the Algorithm~\ref{alg:addAnno} the bag projection is applied before the \texttt{Group By} and creates the column \textit{prov}. Next, Rule~\ref{qy:aggProjGB} is applied to annotate the \texttt{Group By} operation. This ends the annotation of the subquery  and the result shown in  Listing~\ref{qy:exFilter2}.

\begin{lstlisting}[language=SQL, caption={The transformation of the query situated at the lowest level of the traversable tree}, label={qy:exFilter2}]
SELECT sn, |\color{cyan}LISTAGG(CONCAT(prov, '$\otimes$', duration), ' +sum ') WITHIN GROUP (ORDER BY sn) AS total, CONCAT('$\delta$(', LISTAGG(prov, '+') WITHIN GROUP (ORDER BY sn), ')') AS prov|
FROM te_azores
GROUP BY sn
\end{lstlisting}

Then, the main query is annotated by calling the function ``NestedAgg'', which detects a comparison with aggregated values, and applies the Rule~\ref{qy:filter}. The query obtained is in Listing~\ref{qy:exFilter3}. 

\begin{lstlisting}[language=SQL, caption={The transformation of the second level query}, label={qy:exFilter3}]
SELECT C0.sn, C0.total, |\color{cyan}CONCAT(C0.prov, '.[', C0.total, '\textless= 1 $\otimes$ 200]') AS prov|
FROM(
	SELECT sn, |\color{cyan}LISTAGG(CONCAT(prov, '$\otimes$', duration), '+sum') WITHIN GROUP (ORDER BY sn) AS total, CONCAT('$\delta$(', LISTAGG(prov, '+') WITHIN GROUP (ORDER BY sn), ')') AS prov|
	FROM te_azores
	GROUP BY sn) C0
\end{lstlisting}

This query produces the result shown in Table~\ref{tab:filterex2}. Our methods also allow to get only tuples for which the symbolic expression is \textit{True} (Listing 5). To do this, a new column is added with the name of the original column (total, in the example in Listing~\ref{qy:exFilter5}) plus the suffix ``\_agg''. This column ensures that all original comparisons and filters are retained while effectively propagating information for the provenance polynomials based on the new columns.

\begin{lstlisting}[language=SQL, caption={The query~\ref{qy:exFilter} annotated to show only tuples where the symbolic expression results in $1k$}, label={qy:exFilter5}]
SELECT sn, |\color{orange}total, total\_agg|,|\color{cyan}CONCAT(prov, '.[',| |\color{orange}total\_agg||\color{cyan},'\textless= 1 $\otimes$ 200]') AS prov|
FROM(
    SELECT sn, |\color{orange}SUM(duration) total|,  |\color{cyan}LISTAGG(CONCAT(prov, '$\otimes$', duration), '+sum') WITHIN GROUP (ORDER BY sn) AS| |\color{orange}total\_agg||\color{cyan}, CONCAT('$\delta$(', LISTAGG(prov, '+') WITHIN GROUP (ORDER BY sn),')') AS prov|
    FROM te_azores
    GROUP BY sn) C0 |\color{orange}WHERE total \textless= 200|
\end{lstlisting}

\begin{table}
    \centering
    \caption{The result of Listing~\ref{qy:exFilter3}}
   \label{tab:filterex2}
    \begin{tabularx}{\linewidth}{|c|Y|Y|Y|}
      \hline
  \textbf{sn} & \textbf{total}& \textbf{total\_agg} & \textbf{prov} \\ 
\hline
sn123 & 200 & $t1 \otimes 100 +_{sum} t4 \otimes 100$ & $\delta(t1 + t4).[t1 \otimes 100 +_{sum} t4 \otimes 100 <= 1 \otimes 200 ]$ \\
\hline
sn234 & 150 &$t2 \otimes 150$ & $ \delta(t2) . [t2 \otimes 150 <= 1 \otimes 200]$ \\
 \hline
   \end{tabularx}
\end{table}

The result of the query is in Table~\ref{tab:filterex2}, where the two columns in the left are the result of the original query and the columns in the right are the annotations corresponding to the columns $a'_n$ and $prov$ referred in Sections~\ref{sec:SPJUAops} and \ref{sec:ExtendedSemantics}.
The symbolic expressions are preserved so that users can understand how the result was obtained and why those specific tuples contribute to the final result. In this case, the explanation for the tuples that were excluded from the result is omitted. 

\paragraph{Example 2.}
This example demonstrates how to transform the nested query with \texttt{EXISTS} from Listing~\ref{qy:exNested} into SPJUA operations, while also addressing exceptions to ensure the correct computation of provenance annotations.

\begin{lstlisting}[language=SQL, caption={The example of a nested query with Tables~\ref{tab:equip} and \ref{tab:temad}}, label={qy:exNested}]
SELECT e.model 
FROM equipments e 
WHERE EXISTS(
    SELECT model 
    FROM te_madeira tm 
    WHERE e.model <> tm.model) 
GROUP by e.model;
\end{lstlisting}

Applying rule~\ref{qy:exists} in Section~\ref{sec:nested} gives the query illustrated in Listing~\ref{qy:exNested2}. 
However, this transformation introduces an intermediate \texttt{Join} that does not exist in the original query. Capturing its provenance would therefore result in an incorrect provenance polynomial. To address this, we assign the subquery with the \texttt{Distinct} clause the alias ``\texttt{nestedT0}'', and use the alias ``\texttt{nestedT}'' as a flag in our algorithm. This flag signals that when the aggregation function (in this example, \texttt{Listagg}) is applied, the added \texttt{Join} should be excluded from the provenance annotation.

\begin{lstlisting}[language=SQL, caption={The unnesting transformation of the query in Listing~\ref{qy:exNested}}, label={qy:exNested2}]
SELECT e.model, |\color{cyan}LISTAGG(CONCAT(e.prov, ' . ', '(',tm.prov,')'), ' + ') WITHIN GROUP (ORDER BY e.model)|
FROM equipments e |\color{cyan}INNER JOIN|
    |\color{cyan}(SELECT nestedT0.model, LISTAGG(tm.prov, ' + ') WITHIN GROUP (ORDER BY nestedT0.model)| 
    |\color{cyan}FROM te\_madeira tm INNER JOIN (SELECT DISTINCT e.model equipments e) nestedT0 ON e.model \textless \textgreater tm.model)|
    |\color{cyan}GROUP BY nestedT0.model|
|\color{cyan})C0 ON e.model = C0.model| GROUP by e.model
\end{lstlisting}

Finally, the result of the subquery (aliased as ``\texttt{C0}’’) is joined with the outer relation using an equality condition. The outer query is then evaluated by applying the rules for \texttt{Join} and \texttt{Group By} operations (Rules~\ref{qy:join} and~\ref{qy:gp1} from Section~\ref{sec:SPJUAops}). The result of this transformation is presented in Table~\ref{tab:filterex3}.

\begin{table}
    \centering
    \caption{The result of Listing~\ref{qy:exNested2}}
   \label{tab:filterex3}
    \begin{tabularx}{\linewidth}{|c|Y|}
      \hline
  \textbf{model} & \textbf{prov} \\ 
\hline
ModelA & $\delta(t5 . \delta(t8 + t9) + t6 . \delta(t8 + t9))$ \\
\hline
ModelB & $\delta(t7 . \delta(t10))$  \\
 \hline
   \end{tabularx}
\end{table}
\section{Experimental Evaluation}
\label{sec:experimental}

This section provides an evaluation and comparison of DataProv, Perm, GProM, and ProvSQL in terms of expressiveness, query runtime and data provenance size using the TPC-H benchmark~\cite{tpch}.  
It also demonstrates 
how each solution behaves with different database sizes.

\subsection{Environment and Methodology}

The TPC-H benchmark is widely used for evaluating decision-support systems and data warehouse applications. It consists of complex analytical queries involving multiple tables, and complex queries. We used the tools available in TPC-H to generate two databases with sizes 100MB and 1GB and 22 queries.

\begin{figure*}
\centering
    \begin{subfigure}{0.49\textwidth}
    \centering
     \includegraphics[width=\linewidth]{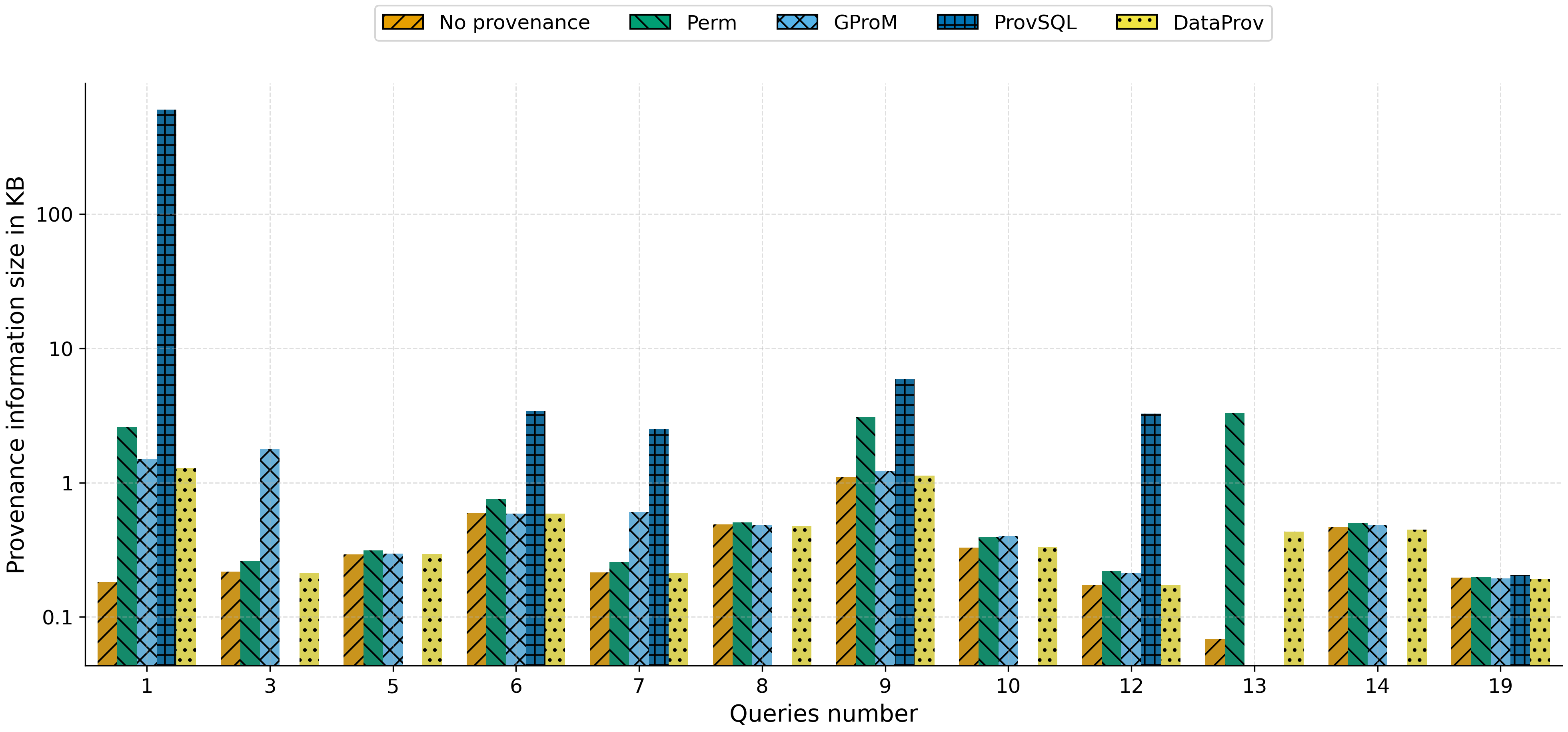}
    \caption{Non-nested queries runtimes for 100MB database}
    \label{fig:nonnested100_run}
    \end{subfigure}
    \begin{subfigure}{0.49\textwidth}
    \centering
     \includegraphics[width=\linewidth]{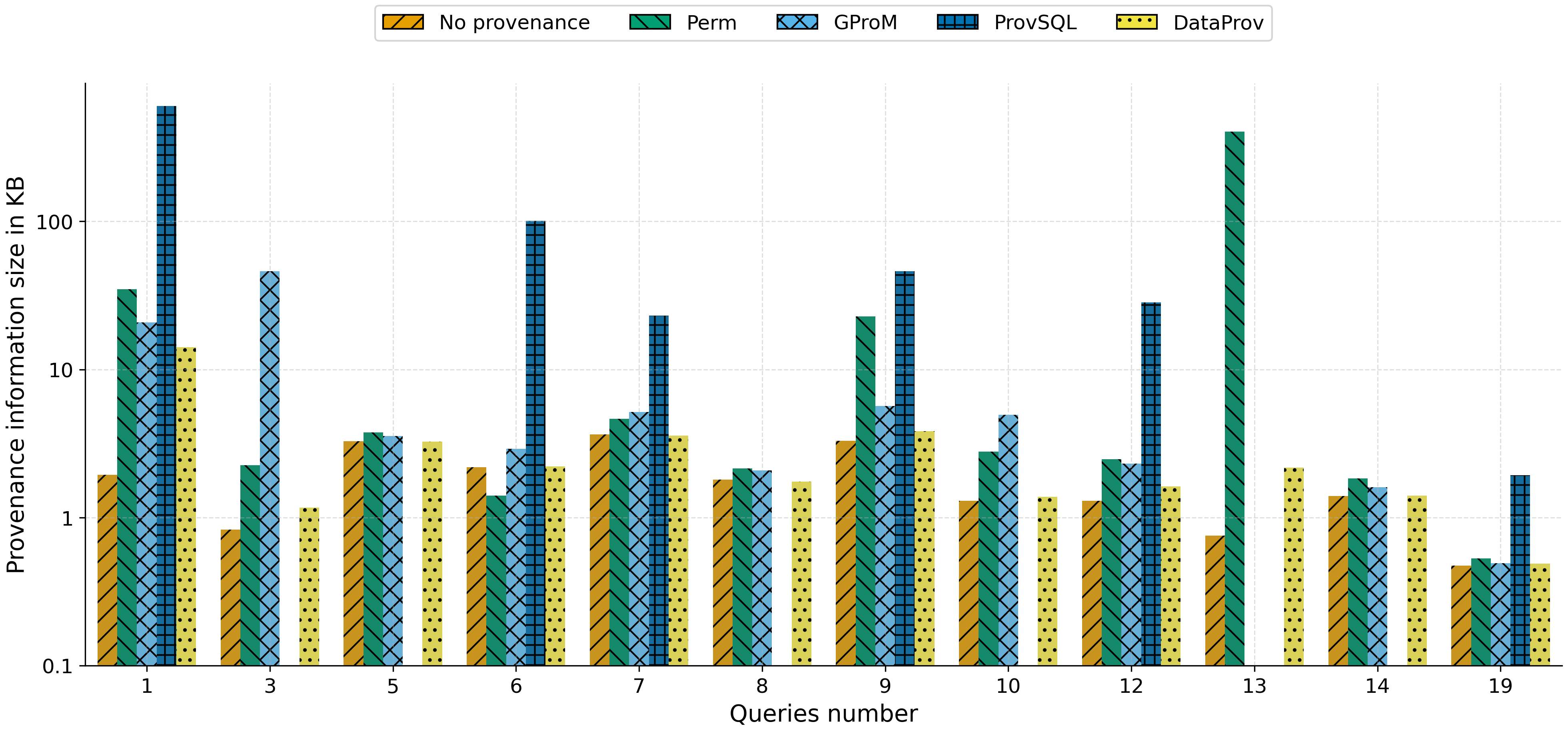}
    \caption{Non-nested queries runtimes for 1GB database}
    \label{fig:nonnested1_run}
    \end{subfigure}
    \begin{subfigure}{0.49\textwidth}
    \centering
     \includegraphics[width=\linewidth]{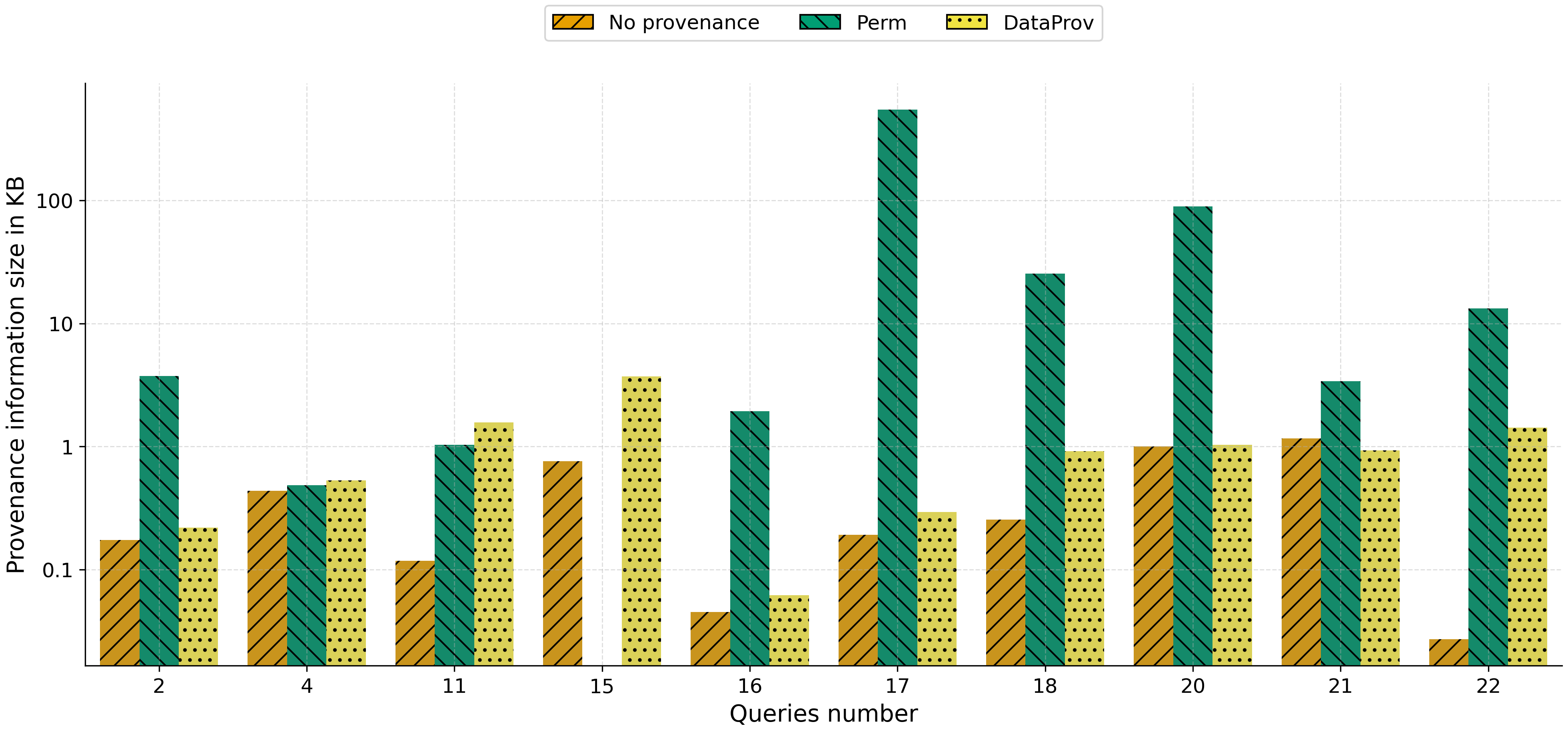}
    \caption{Nested queries runtimes for 100MB database}
    \label{fig:nested100_run}
    \end{subfigure}
    \begin{subfigure}{0.49\textwidth}
    \centering
     \includegraphics[width=\linewidth]{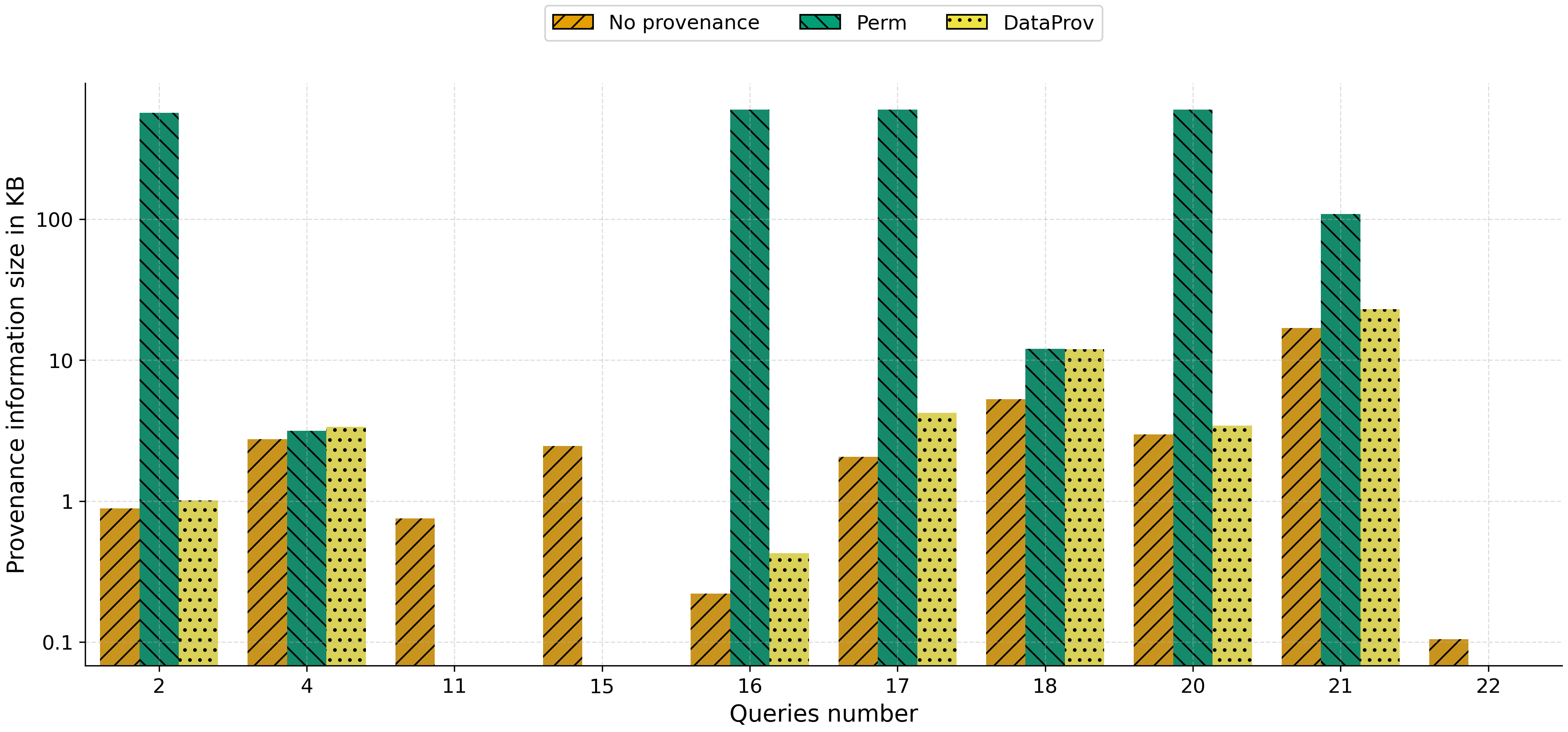}
    \caption{Nested queries runtimes for 1GB database}
    \label{fig:nested1_run}
    \end{subfigure}
    \caption{Queries runtimes results for 100MB and 1GB databases}
    \label{fig:runtimes}
\end{figure*}

The experiments were performed on a machine running Arch Linux (rolling release, updated as of August 2025), equipped with 12 CPUs, 32GB of RAM, and 300GB of disk space. The systems Perm, GProM, and ProvSQL were installed from their respective GitHub repositories, using the latest available versions at the time of testing. All tests were performed on PostgreSQL version 16, as ProvSQL and Perm are tied to this DBMS. 
The primary objective was to evaluate the impact of provenance annotations on query execution time. To ensure a fair comparison, the annotated queries produced by each system were extracted and executed directly on the PostgreSQL database. Additionally, all original queries were unnested to eliminate structural bias and enable a consistent basis for comparison. As previously noted, GProM, unlike Perm, enables users to minimise the provenance columns rather than displaying all columns from every table involved. To ensure a fair comparison of the results, the GProM queries were specifically configured to display only the column containing the provenance tokens (\texttt{prov}) for each table included in the query. Aside from this configuration, the queries were executed using default settings and without any optimisation methods from the frameworks.

In Section~\ref{sec:how_prov_gen}, we introduced two distinct implementations for calculating provenance polynomials based on the extension proposed in~\cite{Yael2011}. To facilitate comparisons with other platforms, we used the implementation that adheres strictly to the method outlined in~\cite{Yael2011} for our method. Following this, it is provided a comparison between our two implementations.

While executing the experiments, we (i) created the tables and indexes, (ii) cleared the database cache before executing every query, (iii) ran every query three times, and (iv) saved the execution times for every query execution. The execution times reported in the following are the averages of the three executions. Queries exceeding a 10-minute time limit were terminated.

\subsection{Validation of results and provenance annotations}
\label{sec:validation}

To ensure the accuracy of our provenance annotations, we used the provenance polynomials validator presented in~\cite{pintor2025}. This tool assesses the validity of provenance polynomials across multiple dimensions. One dimension is the comparison between the output of the original query and the result of the annotated query to confirm that the query result does not change. In cases involving aggregations, it resolves symbolic expressions according to the formalism proposed in~\cite{Yael2011}. This verification process ensures that tuples annotated with $1_K$ correspond to the original result set and that the values in the aggregated columns match those of the originals after the expressions are solved.

The validator also performs structural validation of the provenance annotations. Specifically, it expands the provenance polynomials and examines all annotations involving the \texttt{Join} operator (denoted ``\texttt{.}''). Using the values of the provenance polynomials, it checks if the result of the columns from the rows corresponding to these tokens matches the original. Furthermore, it ensures that all alternative derivations (captured by the ``\texttt{+}'' operator) are correctly included, confirming the completeness of the provenance.

When applying the validator to the provenance polynomials generated by our methods on the TPC-H benchmark queries, all validations succeeded except for Query 15. The failure is due to the presence of a \texttt{Count(Distinct…)} aggregation in the \texttt{Select} clause, which is not supported by our framework. 
To address this issue, we introduce a transformation that rewrites the aggregation as:

\begin{equation}
\begin{gathered}
\myBigGamma_{v,\text{COUNT}(\text{DISTINCT } u)}(R^+)
\Rightarrow
\myBigGamma_{v,\text{COUNT}(*)}(\myBigPi_{v,u}(R^+))
\end{gathered}
\label{qy:countdist}
\end{equation}

\noindent
such that, $v$ denotes the \texttt{Group By} attributes, and $u$ is the column originally used in the \texttt{Distinct} clause. This transformation first projects the relevant columns $(v, u)$ to eliminate duplicates, and then performs an aggregation using \texttt{Count(*)} on the resulting distinct tuples. This rewriting preserves the results of the original query, while ensuring that the provenance polynomials derived using the rules in Section~\ref{sec:notations} are valid.
With this adjustment, all benchmark queries passed the validation step.  

\subsection{Experimental Results}

The results presented in the following are organised into two categories: nested and non-nested queries. The query execution times are in seconds. The Y-axis in all charts is in a logarithmic scale with Base 10. The TPC-H queries are denoted T1, T2, \dots, T22.

\subsubsection{Runtimes comparison}

Figure~\ref{fig:runtimes} shows the runtimes of the 22 queries in the TPC-H benchmark, presented as four distinct charts.  Figure~\ref{fig:nonnested100_run} presents the elapsed times for the non-nested queries in 100MB database. Figure~\ref{fig:nonnested1_run} presents the same for the 1GB database. Figures~\ref{fig:nested100_run} and~\ref{fig:nested1_run} present the elapsed times for the nested queries in the 100MB and 1GB database, respectively.

ProvSQL executed only 6 queries, as it does not support nested queries, \texttt{Order By} on aggregated columns, or complex \texttt{Having} conditions.

GProM executed all queries except T13, which failed due to query rewriting errors, and does not support nested queries. Additionally, execution times for queries T11 and T22 are unavailable in Perm and DataProv with the 1GB dataset, as both systems terminated the process due to out-of-memory exceptions. 
More specifically, in the case of query T22 in DataProv, the issue arises because the provenance annotation results in a string that exceeds PostgreSQL’s 1 GB limit on string size.
DataProv could not also complete T15 on the 1GB dataset due to memory limitations. In Perm, the rewritten version of T15 produces no results, which also prevents the measurement of execution times for both dataset sizes. 

ProvSQL consistently exhibits the highest execution times for queries where it is able to capture provenance, possibly due to the use of cursors within functions and stored procedures.

The execution times for non-nested queries with provenance annotations in the other frameworks are generally comparable to those of the original queries without provenance, indicating a low overhead.
The execution  of Query T13 in Perm, for both database sizes, is an exception.  This is attributed to the Perm's query rewriting strategy, which employs the \texttt{LEFT JOIN} operator to disaggregate the query results and the provenance annotations. 
Upon analysis of the execution plan of the rewritten query, it becomes evident that this operator significantly increases the computational costs.

Figure~\ref{fig:nested100_run} and Figure~\ref{fig:nested1_run}, do not display execution times for GProM because it does not support nested queries~\cite{Niu2019}. This claim is corroborated in GProM's GitHub page~\footnote{\url{https://github.com/IITDBGroup/gprom}}. 

For nested queries, the execution times with provenance can differ greatly from those without provenance.

This is particularly evident for Perm, which exceeded the 600 seconds timeout in multiple queries, even on the 100MB dataset. DataProv demonstrates comparatively better performance and, in some cases, achieves execution times close to those of queries without provenance.

\begin{figure*}
\centering
    \begin{subfigure}{0.49\textwidth}
    \centering
     \includegraphics[width=\linewidth]{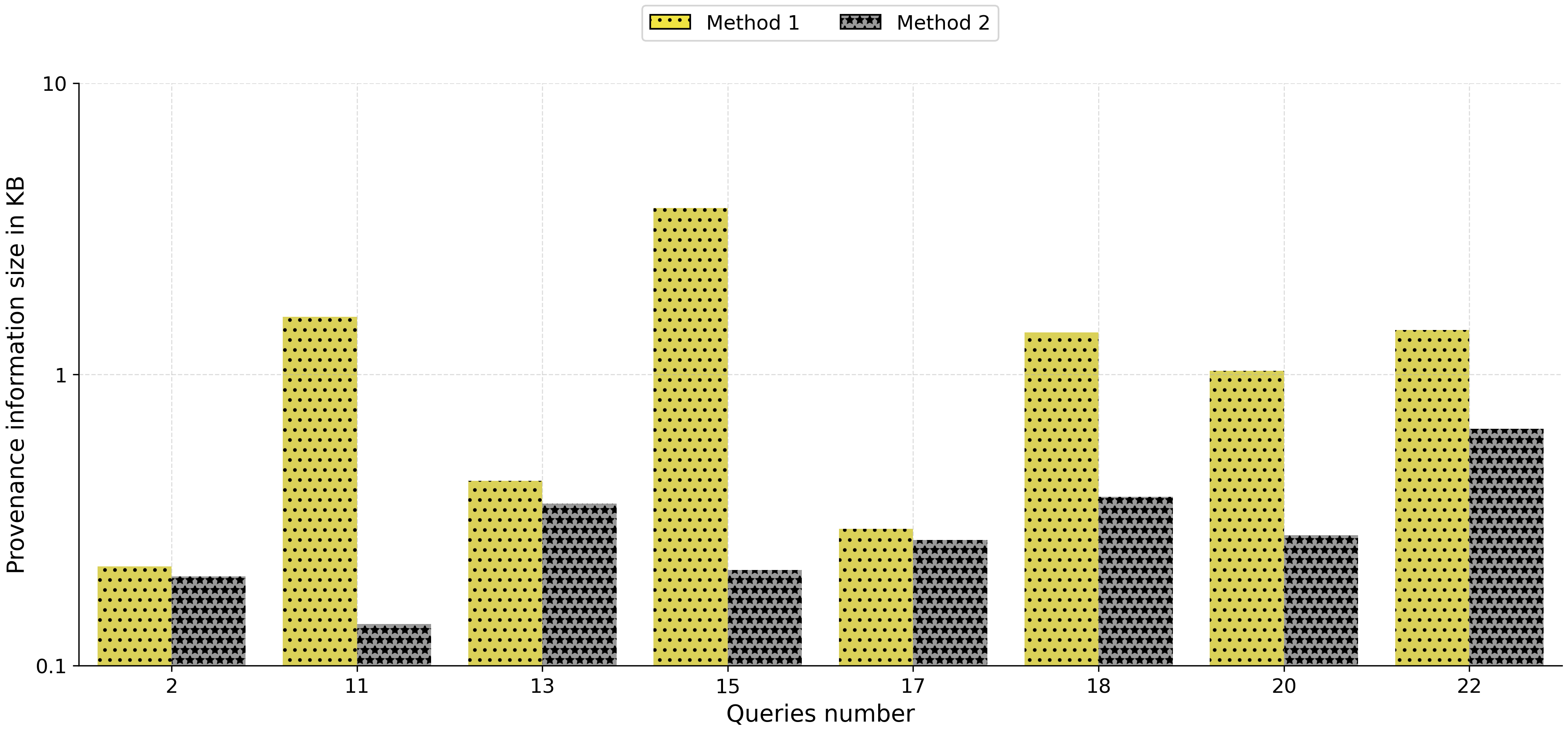}
    \caption{Runtimes comparison between DataProv Method 1 and Method 2 for 100MB database}
    \label{fig:dp100_run}
    \end{subfigure}
    \begin{subfigure}{0.49\textwidth}
    \centering
     \includegraphics[width=\linewidth]{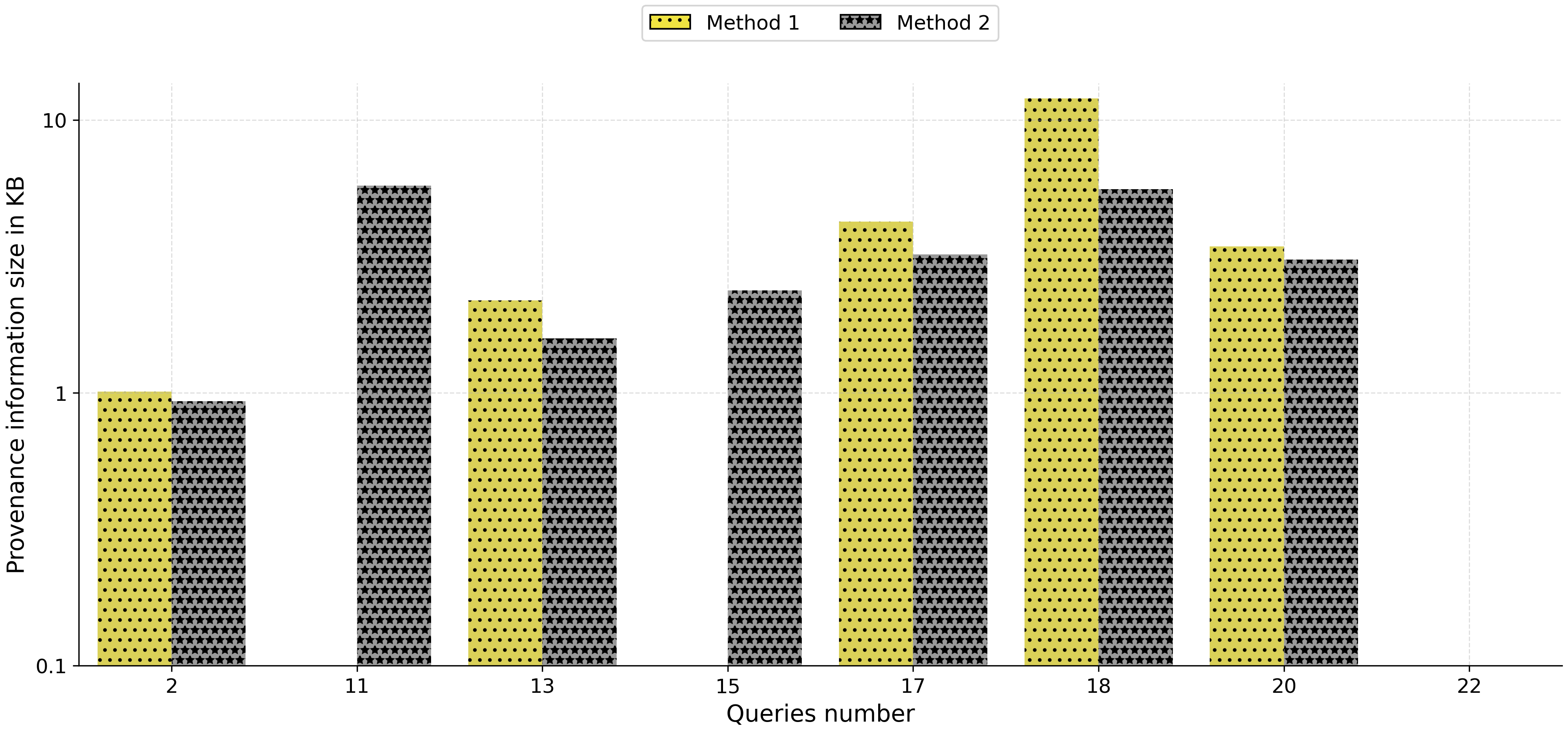}
    \caption{Runtimes comparison between DataProv Method 1 and Method 2 for 1GB database}
    \label{fig:dp1_run}
    \end{subfigure}
    \caption{Queries runtimes for 100MB and 1GB databases comparing our two methods}
    \label{fig:runtimes2}
\end{figure*}

Figure~\ref{fig:runtimes2} presents the runtimes of 8 out of the 22 queries from the TPC-H benchmark, executed on both 100MB and 1GB databases for the two alternatives considered in this work to annotate aggregations. 
These are the queries involving filters or operations over aggregated values, including \texttt{Group By} clauses. For the remaining queries, the only difference introduced is the addition of one or more columns containing the actual values of the aggregate functions, which has a negligible impact on performance and also on the size.

Method~1 corresponds to a full implementation of the approach proposed in~\cite{Yael2011}. Method~2 is the implementation that preserves the symbolic expressions and core principles from~\cite{Yael2011}, but also retains the original aggregate column values, thereby allowing users to recover the original query results directly.
This also means that in selections or joins over aggregations, the results of the aggregations can be used directly in the selection and join conditions to filter out non-qualifying tuples.
As anticipated, Method~2 consistently demonstrates superior runtimes across all queries. Notably, it successfully executes queries T11 and T15 on the 1GB database, which involve comparisons and filters on aggregated values.  In contrast, in Method~1, the aggregated values are only represented as symbolic expressions and so they cannot used directly in comparisons. Thus, the generation of the provenance annotations resorts to performing a Cartesian product between relations, which can significantly impair performance. Yet, Query T22 cannot be executed in both cases due to the size of the generated provenance.

\subsubsection{Data provenance information size}

The size of provenance annotations is $s = t - p$, where $t$ is the size (number of characters) of the query result including the provenance annotations and $p$ is the size of the original query (without provenance). The sizes are  displayed in KB. 

\begin{figure*}
\centering
    \begin{subfigure}{0.49\textwidth}
    \centering
     \includegraphics[width=\linewidth]{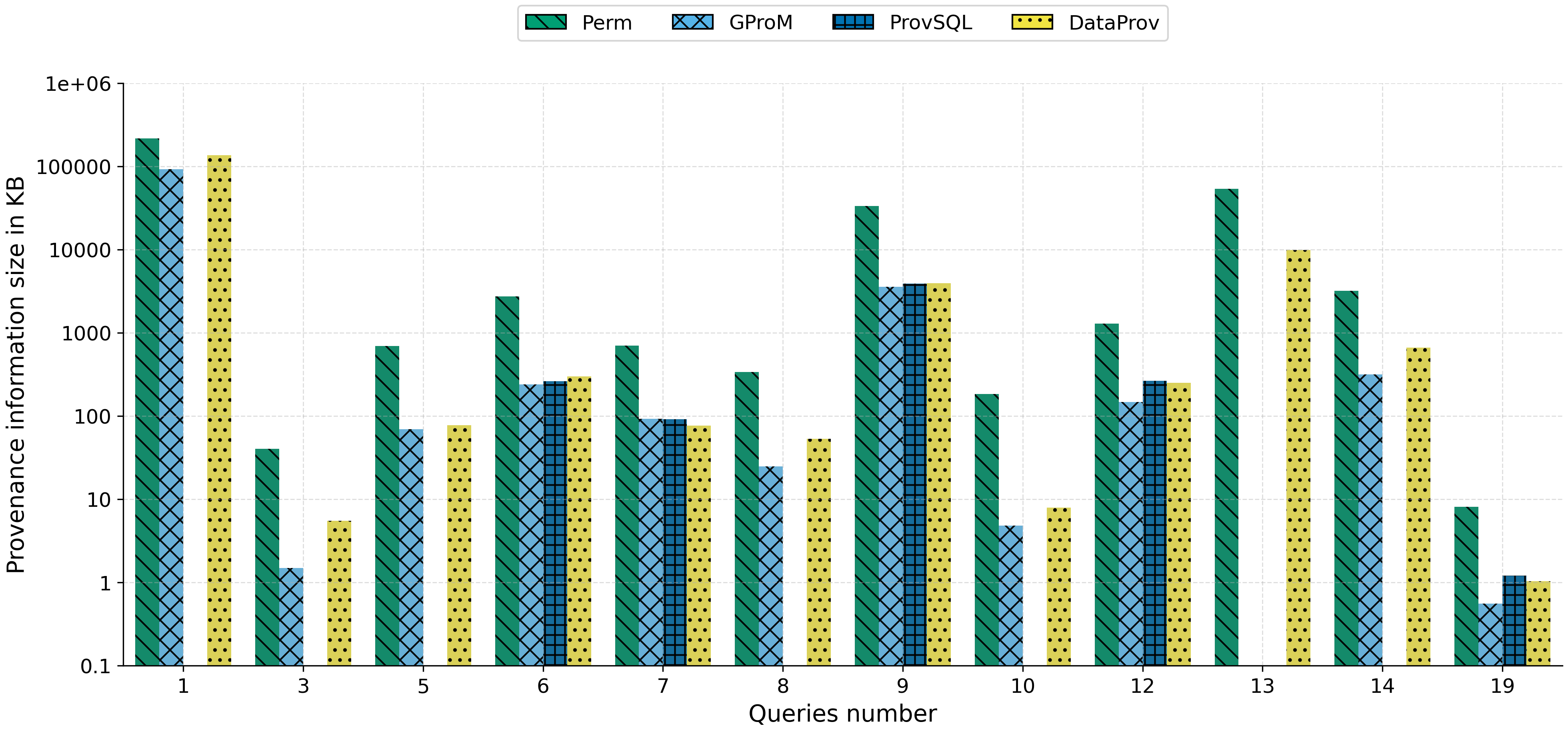}
    \caption{Data provenance information size for non-nested queries in 100MB database}
    \label{fig:nonnested100_size}
    \end{subfigure}
    \begin{subfigure}{0.49\textwidth}
    \centering
     \includegraphics[width=\linewidth]{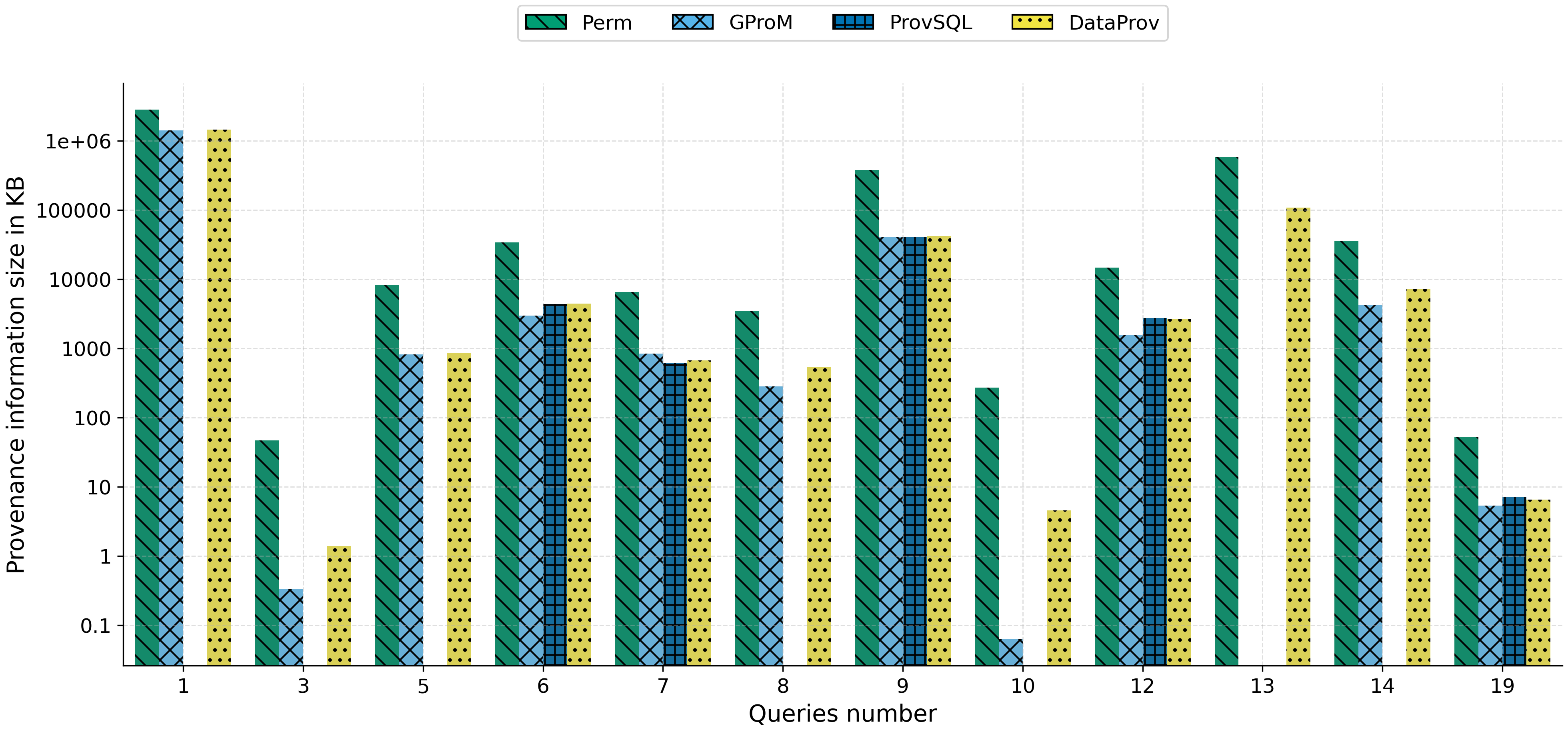}
    \caption{Data provenance information size for non-nested queries in 1GB database}
    \label{fig:nonnested1_size}
    \end{subfigure}
    \begin{subfigure}{0.49\textwidth}
    \centering
     \includegraphics[width=\linewidth]{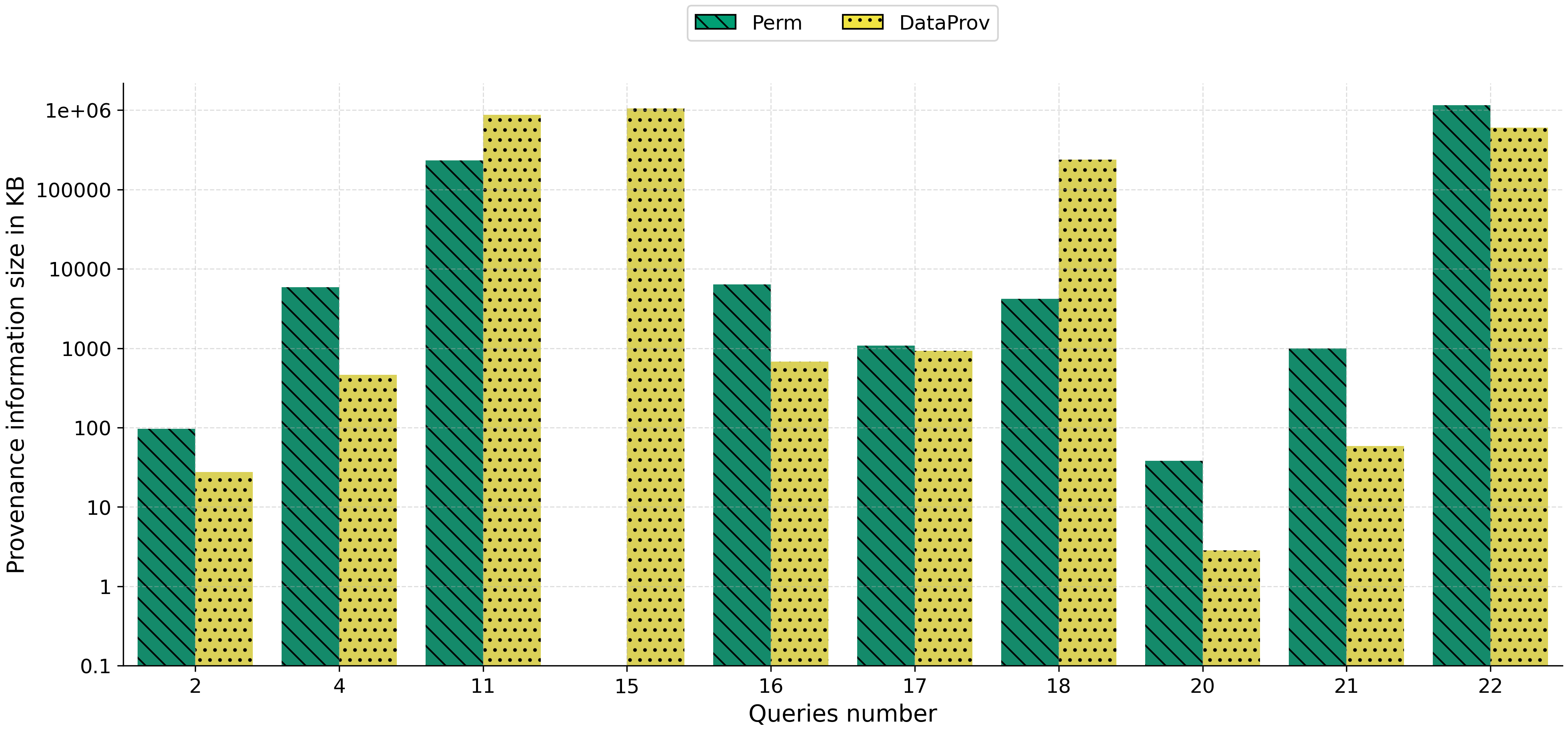}
    \caption{Data provenance information size for nested queries in 100MB database}
    \label{fig:nested100_size}
    \end{subfigure}
    \begin{subfigure}{0.49\textwidth}
    \centering
     \includegraphics[width=\linewidth]{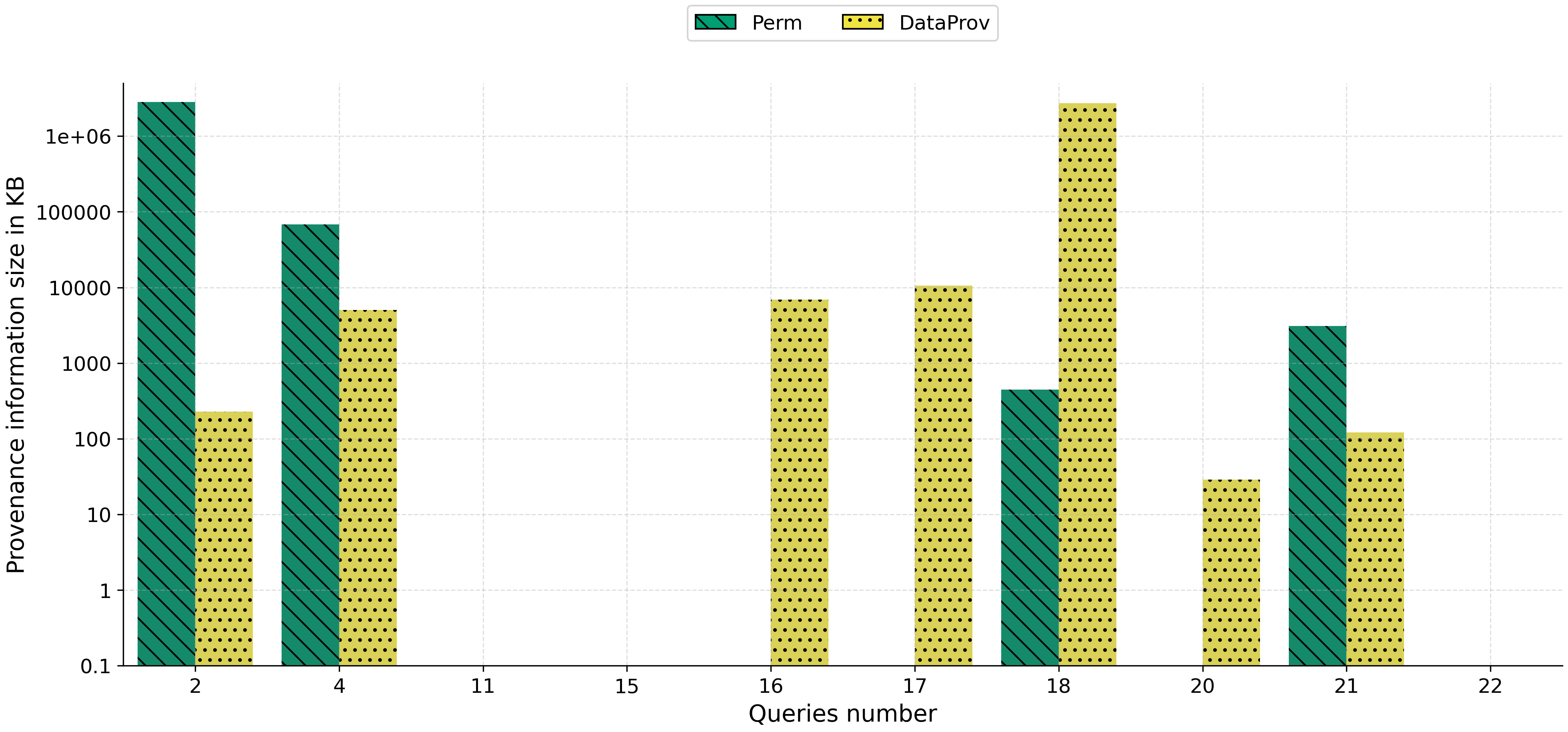}
    \caption{Data provenance information size for nested queries in 1GB database}
    \label{fig:nested1_size}
    \end{subfigure}
    \caption{Data provenance information size for queries in 100MB and 1GB databases}
    \label{fig:size_queries}
\end{figure*}

Figures~\ref{fig:nonnested100_size} and~\ref{fig:nonnested1_size} show the provenance sizes of non-nested queries for the 100MB and 1GB databases, respectively, while Figures~\ref{fig:nested100_size} and~\ref{fig:nested1_size} show the sizes for nested queries on the same datasets.
The queries with no results either exceeded the specified time limit or represent cases where no values were available, as previously discussed during the presentation of execution times.

Perm consistently produces the largest provenance sizes across non-nested queries. GProM’s ability to reduce the number of provenance columns has a noticeable impact, resulting in significantly smaller sizes compared to Perm.
When compared to DataProv, the sizes are very similar in most cases, although GProM often exhibits a smaller size.

When comparing DataProv and ProvSQL, the provenance sizes are consistently similar, with only minor differences. This similarity stems from the fact that both frameworks represent provenance in a comparable manner, with variations mainly resulting from slight differences in the annotation format used for handling aggregation queries following the approach in~\cite{Yael2011}.

For nested queries, Perm generally produces larger provenance sizes than DataProv, with a few exceptions, most notably in Queries T11 and T18, both of which include filters over aggregated values. Interestingly, for query T18, the 1GB database produces fewer result tuples than the 100MB database due to the applied filter, resulting in a smaller provenance size for Perm on the larger dataset.

\begin{figure*}
\centering
    \begin{subfigure}{0.49\textwidth}
    \centering
     \includegraphics[width=\linewidth]{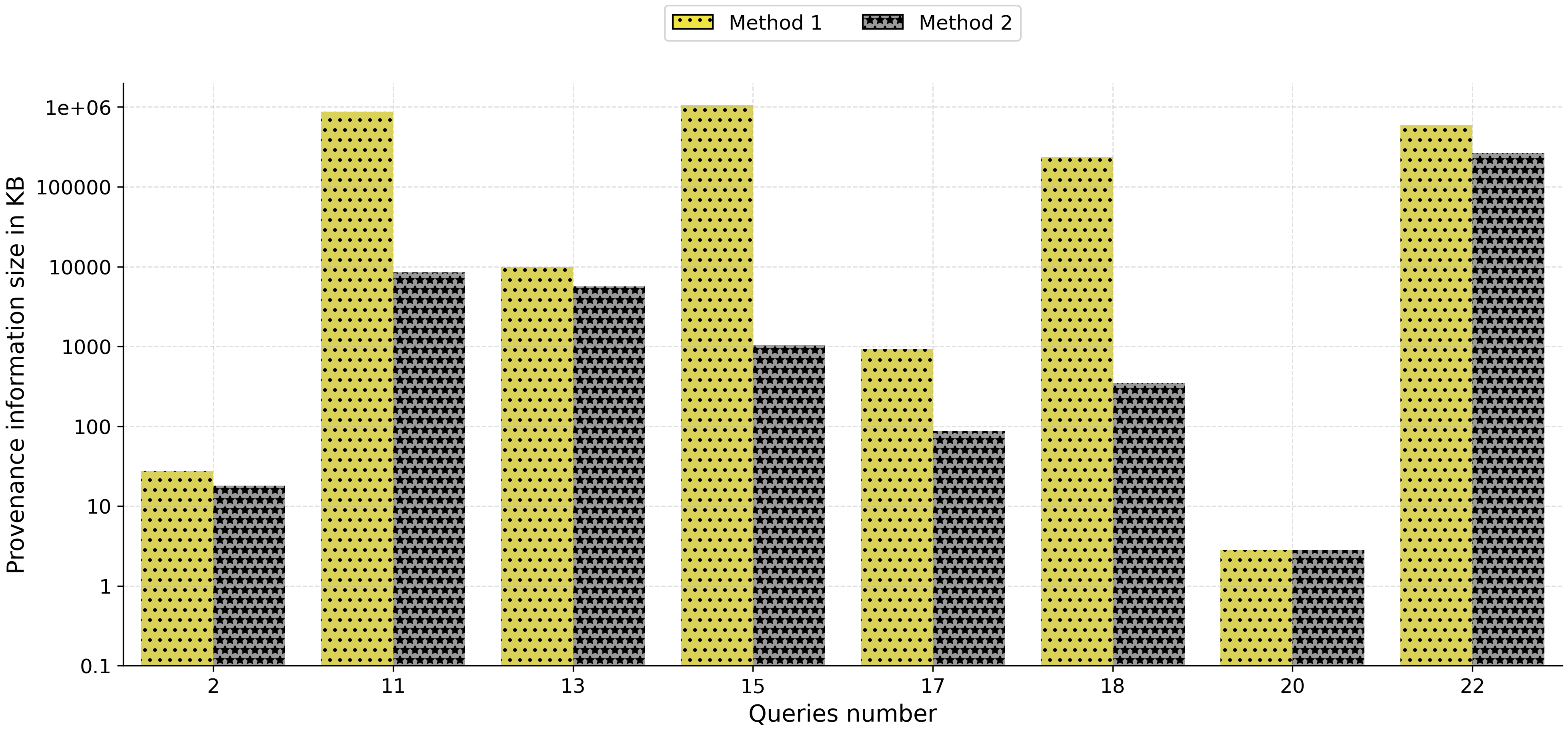}
    \caption{Information size comparison between DataProv Method 1 and Method 2 for 100MB database}
    \label{fig:dp100_size}
    \end{subfigure}
    \begin{subfigure}{0.49\textwidth}
    \centering
     \includegraphics[width=\linewidth]{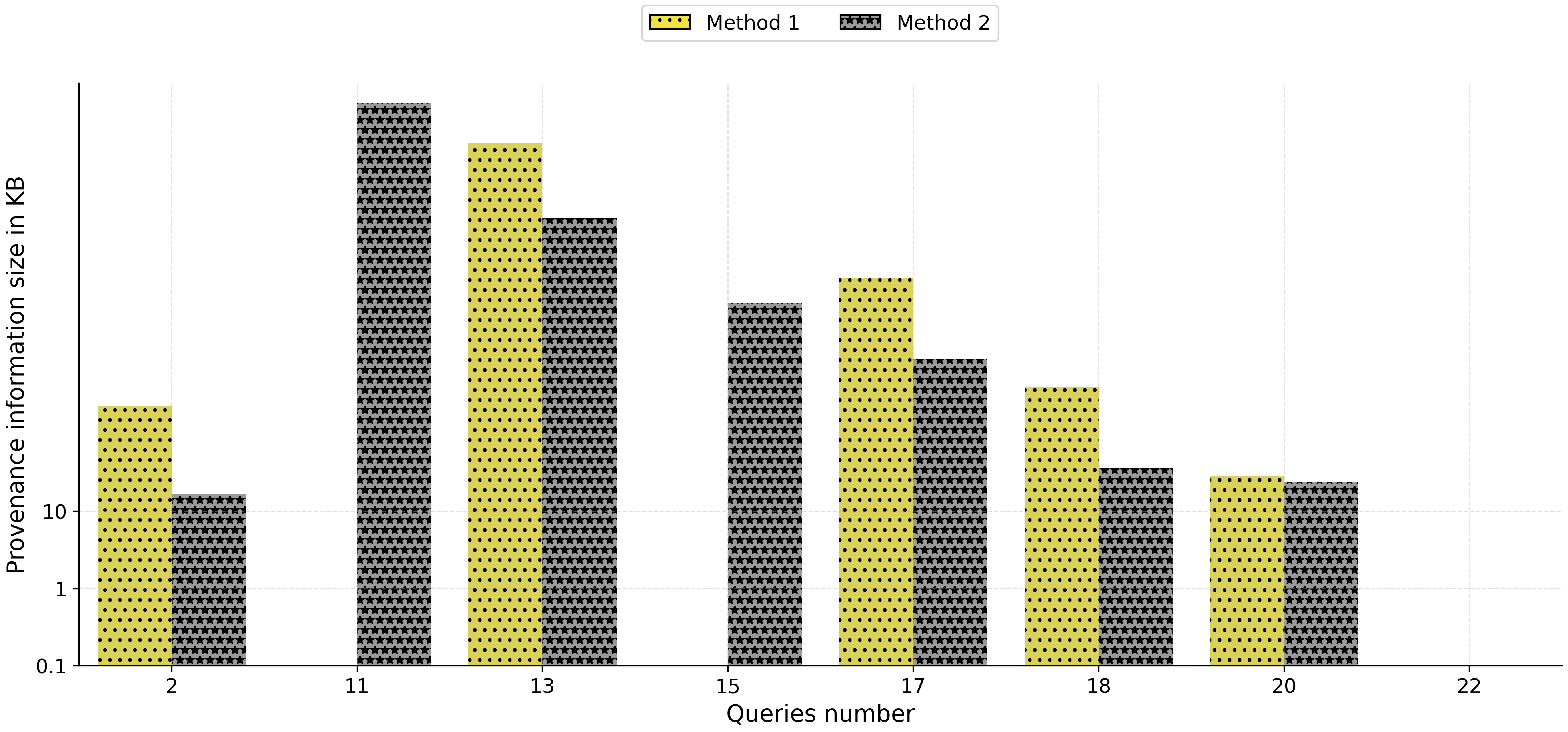}
    \caption{Information size between DataProv Method 1 and Method 2 for 100MB database}
    \label{fig:dp1_size}
    \end{subfigure}
    \caption{Data provenance information size for queries in 100MB and 1GB databases comparing our two methods}
    \label{fig:size_queries2}
\end{figure*}

Figure~\ref{fig:size_queries2} presents the size of the provenance annotations produced by Methods 1 and 2 across different database sizes. Similar to the runtime results, Method~2 consistently shows better performance. This improvement is particularly evident in Queries T11, T15, and T18 in 100MB database, where comparisons are performed over aggregated columns.

It is worth noting that Query T20 yields similar provenance sizes in both methods, as the comparison condition has minimal influence on the final result. Additionally, query T11 in the 1GB database under Method~2 produces the largest provenance size among all queries. 
This explains why the same query cannot be executed under Method~1: the annotations generated by Method 2, which only outputs the tuples annotated with $1_{K}$ already results in a very large output, including all tuples, either annotated with  $1_{K}$ and $0_{K}$, as in Method~1, further increases the result size.

In contrast, Query T15 shows a more moderate increase in provenance size, in 1GB database under Method~2, yet it still cannot be executed using Method~1, again demonstrating the significant overhead introduced by comparisons over aggregated values.

\subsection{Discussion}

Perm, GProm, ProvSQL and DataProv can be categorised in different ways, regarding the architecture, the proposed methods and the results they provide. Perm, GProm and DataProv use query rewriting techniques to build data provenance using only standard SQL operations. ProvSQL uses user-defined functions and cursors implemented in PL/pgSQL with provenance circuits. 

On the other hand, ProvSQL and DataProv align closely with the formalisms proposed in~\cite{Green2007,Yael2011}. Their results are expressed as provenance polynomials, from which other simpler algebraic structures can be derived. The outputs of the original queries remain unchanged, except in the presence of aggregations, which are handled following the methods outlined in~\cite{Yael2011}. In contrast, Perm and GProM disaggregate provenance annotations into multiple tuples and do not provide explicit information on \texttt{Aggregation} operations, as proposed in~\cite{Yael2011}.

GProM and DataProv are middleware solutions. However, GProM does not currently support nested queries in the \texttt{Where} clause. Consequently, the comparison of execution times was limited to non-nested queries. Both Perm and ProvSQL are built on top of PostgreSQL.

Table~\ref{tab:my_label} presents a summary of the overall results for the four frameworks using 100MB and 1GB databases. The  columns \textit{Successfully executed} indicate the  number of TPC-H queries that each framework was able to execute, excluding those that exceeded the 10-minute threshold. DataProv stands out as the framework with the highest expressive power. The columns \textit{smallest execution times} list how often each framework achieved the best execution times. Once again, DataProv demonstrated strong performance, achieving the fastest execution in the most significant number of queries across both database sizes.

\begin{table}
\centering
\caption{An overall summary of the results for the four frameworks in the 100MB and 1GB databases - In number of queries.}
\label{tab:my_label}
\vspace{-0.1cm}
\begin{tabular}{|c|c|c|c|c|c|c|}
\hline
\multicolumn{1}{|c|}{} & \multicolumn{2}{c|}{\textbf{Successfully}} & \multicolumn{2}{c|}{\textbf{Smallest}} & \multicolumn{2}{c|}{\textbf{Smallest}} \\ 
\multicolumn{1}{|c|}{} & \multicolumn{2}{c|}{\textbf{executed}} & \multicolumn{2}{c|}{\textbf{execution}} & \multicolumn{2}{c|}{\textbf{resultset}} \\ 
\multicolumn{1}{|c|}{} & \multicolumn{2}{c|}{} & \multicolumn{2}{c|}{\textbf{time}} & \multicolumn{2}{c|}{\textbf{size}} \\ 
\cline{2-7}
 & \textbf{\textit{100\tiny{MB}}} & \textbf{\textit{1\tiny{GB}}} & \textbf{\textit{100\tiny{MB}}} & \textbf{\textit{1\tiny{GB}}} & \textbf{\textit{100\tiny{MB}}} & \textbf{\textit{1\tiny{GB}}} \\ 
\hline
ProvSQL  &  5 &  5 & 0 & 0 &  1 &  1 \\ 
\hline
GProM    & 11 & 11 & 0 & 0 & 9 & \textbf{10} \\ 
\hline
Perm     & 21 & 17 & 0 & 2 &  2 &  1 \\ 
\hline
DataProv & \textbf{22} & \textbf{19} & \textbf{22} & \textbf{18} & \textbf{10} & 7 \\ 
\hline
\end{tabular}
\end{table}

The last two columns, regarding the size of provenance annotations, DataProv produced the smallest annotations for more queries in the 100MB database. However, in the 1GB setting, GProM yielded the smallest provenance sizes in 10 out of the 11 queries it was able to execute.

Tables~\ref{tab:100mbtime} and~\ref{tab:1gbtim} list the overall execution times of the TPC-H queries for the four frameworks using 100MB and 1GB datasets. The columns represent the TPC-H query IDs and the values indicate the overhead associated with computing provenance annotations relative to the execution times of the original queries (without provenance). 
\begin{table}
\centering
\caption{The average execution overhead (in \%) for queries in each column for the 100Mb database.
    }
    \label{tab:100mbtime}
    \vspace{-0.1cm}
\begin{tabular}{|l|rccc|}
\hline
         & \multicolumn{4}{c|}{\textbf{TPC-H Queries}}                                                                                                                \\ \cline{2-5} 
         & \multicolumn{1}{c|}{1, 6, 7} & \multicolumn{1}{c|}{3, 5, 8}       & \multicolumn{1}{c|}{\small{2, 4, 11, 13, 16}}  &                                 \\
         & \multicolumn{1}{c|}{9, 12, 19}  & \multicolumn{1}{c|}{10,	14}       & \multicolumn{1}{c|}{\small{17, 18, 20, 21, 22}} & 15  \\ \hline
ProvSQL  & \multicolumn{1}{r|}{24848\%}      & \multicolumn{1}{c|}{-}           & \multicolumn{1}{c|}{-}                  & -                                 \\ \hline
GProM    & \multicolumn{1}{r|}{75\%}          & \multicolumn{1}{r|}{97\%} & \multicolumn{1}{c|}{-}                  & -                                 \\ \hline
Perm     & \multicolumn{1}{r|}{187\%} & \multicolumn{1}{r|}{12\%}          & \multicolumn{1}{r|}{19672\%}             & \textbf{-}                        \\ \hline
DataProv & \multicolumn{1}{r|}{\textbf{13\%}} & \multicolumn{1}{r|}{\textbf{0\%}}          & \multicolumn{1}{r|}{\textbf{122\%}}     & \multicolumn{1}{r|}{\textbf{391\%}} \\ \hline
\end{tabular}
\end{table}

\begin{table}
\centering
\caption{The average execution overhead (in \%) for the queries in each group on the 1GB database.}
\label{tab:1gbtim}
\begin{tabularx}{\linewidth}{|Y|Y|Y|Y|}
\hline
         & \multicolumn{3}{c|}{\textbf{TPC-H Queries}} \\ \cline{2-4}
         & 1, 6, 7, 9, 12, 19 & 3, 5, 8, 10, 14 & \small{2, 4, 13, 16, 17, 18, 20, 21} \\ \hline
ProvSQL  & 6126\%            & -               & -                            \\ \hline
GProM    & 191\%             & 577\%           & -                            \\ \hline
Perm     & 420\%             & 49\%            & 8728\%                        \\ \hline
DataProv & \textbf{101\%}    & \textbf{4\%}    & \textbf{51\%}                \\ \hline
\end{tabularx}
\end{table}

These results are grouped into disjoint sets of queries, since not all frameworks run the same queries. 

For the 100MB database, ProvSQL consistently exhibits the highest overhead among all frameworks for the queries it is able to execute, whereas DataProv demonstrates the lowest runtime impact. In the subset of queries where all the frameworks are able to run, GProM outperforms Perm in terms of overhead. However, in the group of queries executable only by GProM, Perm, and DataProv, Perm achieves better performance than GProM. In Table~\ref{tab:100mbtime}, the second column shows DataProv with a 0\% overhead due to rounding, as it is marginally slower than the corresponding query without provenance.

For the 1GB database, the trend remains consistent. DataProv continues to show the lowest overhead across the three groups of queries. Similarly to the 100MB case, GProM outperforms Perm in the queries shared with ProvSQL, while Perm again achieves lower overhead than GProM in the queries that exclude ProvSQL.

We also assessed the impact of query rewriting on the overall execution times of DataProv. The average was just 0.32 seconds per query. 
We were unable to do the same for ProvSQL, Perm and GProM, but we believe the costs would be similar.

Tables~\ref{tab:100mbsize} ands~\ref{tab:1GBsize} list the overall provenance annotation sizes of the TPC-H queries for the four frameworks using 100MB and 1GB databases. 
The results are also grouped into distinct sets of queries. However, these sets differ from the ones used to compare the runtimes because some queries exceeded the 600-second execution limit, which means that provenance sizes are unavailable for those cases.

The overhead introduced by provenance annotations increases significantly for queries that aggregate a large number of tuples. Generally, this overhead is more pronounced for Perm. This is not only because it constructs provenance annotations by including the values of the tuples that contribute to the query results, but also because it requires the replication of rows in the presence of aggregations, while it does not allow for column reduction like GProM.

\begin{table}[ht!]
\centering
\caption{The average provenance size overhead (in \%) for queries in each column for the 100MB database.
        }
    \label{tab:100mbsize}

\begin{tabularx}{\linewidth}{|X|Y|Y|Y|Y|}
\hline
         & \multicolumn{4}{c|}{\textbf{TPC-H Queries}}                         \\ \cline{2-5} 
         & 6, 7, 9, 12, 19 & 1, 3, 5, 8, 10, 14 & \small{2, 4, 11, 13, 16, 17, 18, 20, 21, 22} & 15                                 \\ \hline
ProvSQL  & {$1,95\times10^3\%$}          &  -                 & -                   & -                                      \\ \hline
GProM    & $\bm{1,76\times10^3\%}$          & $\bm{4,68\times10^4\%}$          & -                & -                                      \\ \hline
Perm     & $1,66\times10^4\%$ & $1,11\times10^7\%$ & $\bm{3,72\times10^5\%}$   & -                            \\ \hline
DataProv & $1,97\times10^3\%$ & $6,85\times10^4\%$ & $4,41\times10^5\%$ & $\bm{1,40\times10^6\%}$ \\ \hline
\end{tabularx}
\end{table}

\begin{table}[ht!]
\centering
\caption{The average provenance size overhead (in \%) for queries in each column for the 1GB database. 
        }
    \label{tab:1GBsize}

\begin{tabularx}{\linewidth}{|X|X|X|X|X|}
\hline
         & \multicolumn{4}{c|}{\textbf{TPC-H Queries}}                         \\ \cline{2-5} 
         & 6, 7, 9, 12, 19 & 1, 3, 5, 8, 10, 14 & 2, 4, 13, 18, 21 & 16, 17, 20                                \\ \hline
ProvSQL  & {$3,67\times10^3\%$}          &  -                 & -                   & -                                      \\ \hline
GProM    & $\bm{3,47\times10^3\%}$          & $\bm{2,11\times10^5\%}$          & -                & -                                      \\ \hline
Perm     & $3,25\times10^3$ & $1,11\times10^7\%$ & $1,25\times10^5\%$   & -                            \\ \hline
DataProv & $3,75\times10^3\%$ & $2,17\times10^5\%\%$ & $\bm{1,02\times10^5\%}$ & $\bm{2,02\times10^1\%}$ \\ \hline
\end{tabularx}
\end{table}

For the 100MB database, Perm exhibits the highest overhead in terms of size, even when compared to ProvSQL. Among the queries it executes, GProM consistently demonstrates the smallest impact. When comparing DataProv solely with Perm, Perm shows a significantly lower overhead in 100MB (Table~\ref{tab:100mbsize}).

In the 1GB database, GProM maintains the lowest impact among the queries it executes, and DataProv shows significantly lower overhead for queries executed exclusively by this framework and Perm. It is worth noting that, for the 1GB database, no framework was able to execute queries T11, T15 and T22.

Additionally, in the 100MB database, there is one query for which only DataProv was able to provide results, as the other frameworks either exceeded the 10-minute runtime limit or encountered errors during query rewriting that prevented result generation. In the 1GB database, this number increases to three queries.

\begin{table}
    \centering
    \caption{Average overhead (in \%) comparison between Method 1 and Method 2 from our solution}
    \label{tab:runtime_comparison}
    \begin{tabularx}{\linewidth}{|Y|Y|Y|Y|Y|}
        \hline
        & \multicolumn{2}{c|}{\textbf{100MB}} & \multicolumn{2}{c|}{\textbf{1GB}} \\
        \cline{2-5} 
        & \textbf{Runtime} & \textbf{Size} & \textbf{Runtime} & \textbf{Size} \\
        \hline
        \textbf{Method 1} & 287\% &$1,04\times10^6\%$ & 91\% & $1,64\times10^3$\% \\
        \hline
        \textbf{Method 2} & 0\%  & $1.53\times10^4\%$ & 20\% & $1,06\times10^5$\% \\
        \hline        
    \end{tabularx}
\end{table}

Table~\ref{tab:runtime_comparison} presents a comparison of the two methods in terms of average overhead, considering both runtime and result size.
The results confirm the trends already in Figures~\ref{fig:dp100_run}, \ref{fig:dp100_size}, \ref{fig:dp1_run}, and \ref{fig:dp1_size}, where Method~2 consistently outperforms Method~1 across both metrics. The most significant difference lies in the provenance size: Method~1 produces substantially larger results, as it includes not only the tuples that satisfy the original query but also those that could potentially contribute to it, as the symbolic expressions are not evaluated beforehand. 
The reported 0\% overhead in Method~2 means that the overhead is negligible.

The relative difference in result size is smaller for the 1GB dataset compared to the 100MB dataset  due to the exclusion of three queries — T11, T15, and T22 — from the 1GB results.
There are no results for queries T11 and T15 in Method~1 and for T22 in both methods, both due to memory constraints.

The choice between methods should be guided by the user’s objectives: Method~2 provides more informative and interpretable outputs, whereas Method~1 offers a complete representation, albeit with greater overhead.

\section{Conclusions and Future Work}
\label{sec:conclusion}

This paper introduces methods for generating provenance polynomials using annotations, rewriting rules, and transformations based on standard SQL. These methods can handle multiple operators and nested queries, and are DBMS-independent. Users do not need to modify the original query to obtain provenance polynomial results.

These are the first methods to provide a complete implementation of the approach proposed in~\cite{Yael2011}, which extends semiring theory~\cite{Green2007} to handle aggregations. Additionally, users can choose different types of results, depending on their analytical needs. This work addresses a key limitation of prior research, which either lacked implementation or provided only partial support for the theoretical proposals found in the literature. By leveraging distributed query engines, our approach can be applied in distributed environments, including NoSQL databases.

We implemented our methods in DataProv, a solution that is DBMS-independent, and performed the first comparison of data provenance frameworks. We used a well-known benchmark to demonstrate the expressiveness, performance, and scalability of our methods and other solutions in the literature. Our methods successfully executed queries that were not supported by other platforms, achieved the best performance in terms of execution time overhead, and demonstrated promising results regarding size overhead. 
In some queries, GProM was the best-performing platform in terms of size overhead, although it presents the results differently from our approach. On the other hand, our solution produces well-formed polynomials that fully comply with the formalisms proposed in~\cite{Green2007, Yael2011}. 

The results also indicate that provenance annotations can significantly increase the size of query results, particularly in the presence of aggregations, which may lead to memory issues or output 
truncation due to length limitations. This highlights the need for further research into techniques for summarising and approximating provenance annotations. While not a new topic, it has received limited attention in previous work.

In future work, we plan to explore non-monotone queries, which require the use of the monus operator. Additionally, we aim to investigate post-processing techniques for our results to support a wider range of analyses and visualisations, including the generation of provenance circuits.

\section*{Acknowledgements}
Paulo Pintor has a research grant awarded by FCT - Foundation for Science and Technology - the Portuguese public agency for science, technology and innovation, under the reference - 2021.06773.BD. This work is partially funded by National Funds through the FCT in the context of the projects UIDB/04524/2020 and UIDB/00127/2020, and under the Scientific Employment Stimulus - Institutional Call - CEECINST/00051/2018.

\bibliographystyle{elsarticle-num} 
\bibliography{elsarticle-template-num}

\end{document}